\documentclass[11pt]{article} 
\usepackage{amssymb} 
\usepackage{epsfig} 
\newlength{\bredde} 
\def\slash#1{\settowidth{\bredde}{$#1$}\ifmmode\,\raisebox{.15ex}{/} 
\hspace*{-\bredde} #1\else$\,\raisebox{.15ex}{/}\hspace*{-\bredde} #1$\fi} 
\newcommand{\be}{\begin{equation}} 
\newcommand{\ee}{\end{equation}} 
\newcommand{\bea}{\begin{eqnarray}} 
\newcommand{\eea}{\end{eqnarray}} 
\newcommand{\nn}{\nonumber} 
\newcommand{\al}{\alpha} 
\newcommand{\la}{\lambda}

\newcommand{\om}{\omega} 
 
\newcommand{\Dirac}{\rlap{\hspace{-.6mm} \slash} D} 
\newcommand{\one}{\mbox{\bf 1}} 
\newcommand{\qq}{\langle \bar{q}q\rangle}

\newcommand{\sect}[1]{\setcounter{equation}{0}\section{#1}}

\def\Tr{{\mbox{Tr}}} 
 
\def\im{{\Im\mbox{m}}}

\begin{document} 
\topmargin -1.4cm 
\oddsidemargin -0.8cm 
\evensidemargin -0.8cm 
\title{\Large{{\bf  
On matrix model partition functions for QCD with chemical potential  
}}} 
 
\vspace{1.5cm} 
\author{~\\{\sc G. Akemann}$^1$, {\sc Y.V. Fyodorov}$^{2,3}$ 
and {\sc G. Vernizzi}$^{1}$\\~\\ 
$^1$Service de Physique Th\'eorique, CEA/DSM/SPh$\!$T Saclay\\ 
Unit\'e associ\'ee CNRS/SPM/URA 2306\\ 
F-91191 Gif-sur-Yvette Cedex, France\\~\\ 
$^2$Department of Mathematical Sciences, Brunel University\\ 
Uxbridge, UB8 3PH, United Kingdom\\~\\ 
$^3$ Petersburg Nuclear Physics Institute \\ 
Gatchina 188350, Russia}

\maketitle 
\vfill 
\begin{abstract} 
Partition functions of two different matrix models for QCD with 
chemical potential are computed for an arbitrary number of quark and 
complex conjugate anti-quark flavors. In the large-$N$ limit of weak 
nonhermiticity complete agreement is found between the two 
models. This supports the universality of such fermionic partition 
functions, that is of products of characteristic polynomials in the 
complex plane.   
In the strong nonhermiticity limit agreement is found for an equal  
number of quark and conjugate flavours. 
For a general flavor content the equality of partition functions  
holds only for small chemical potential. 
The chiral phase transition is analyzed for an arbitrary number of quarks,  
where the free energy presents a discontinuity of 
first order at a critical chemical potential. 
In the case of nondegenerate flavors there is first order phase 
transition for each separate mass scale. 
\end{abstract} 
\vfill

\begin{flushleft} 
SPhT T04/029 
\end{flushleft} 
\thispagestyle{empty} 
\newpage

\renewcommand{\thefootnote}{\arabic{footnote}} 
\setcounter{footnote}{0}

\sect{Introduction}\label{intro}

The idea to use random matrix models as a simple model to study the 
non-perturbative phenomenon of chiral symmetry breaking in QCD 
\cite{SV93} has been very successful. Apart from direct studies 
through lattice QCD or other effective models, it has become one of 
the available tools in this area, and we refer to \cite{VW} for a 
review on the matrix model approach.  In the broken phase at zero 
temperature and chemical potential the applicability of a matrix model 
has been completely understood by rederiving part of its results from 
the underlying effective chiral Lagrangian picture, that describes the 
pseudo-Goldstone fields. This has been achieved for the partition 
function \cite{JSV}, the spectral correlation functions of Dirac 
operator eigenvalues \cite{DOTV,TV1,SV03}, as well as for individual 
eigenvalue correlations \cite{ADp}. 
 
A random matrix model that includes the effect of a baryonic chemical 
potential was introduced in \cite{Steph}, studying the nature of the 
quenched approximation and the chiral phase transition. The same model 
was enlarged to include the effect of temperature, and the phase 
diagram of QCD with two light flavors was predicted \cite{HJSSV}, 
including the existence of a tricritical point.  The analysis was 
repeated very recently in \cite{KTV}, distinguishing between baryon 
and isospin chemical potential. Only the matrix model with isospin 
chemical potential can be related to an effective chiral Lagrangian of 
quenched QCD \cite{TV} so far. 
 
The unquenched matrix model partition function with chemical potential for  
a single flavor has been 
analyzed in great detail, including its analytic solution and the behavior of 
its zeros \cite{HJV,HOSV}.  
It has been used as a test case in \cite{HOSV,AANV} for  
lattice algorithms for chemical potential.  
A complete and detailed solution of the unquenched matrix model partition  
function with several, non degenerate flavors has been lacking so far, apart 
from a first attempt in \cite{A03}. It is one of the 
purposes of this article to provide such a solution, both for finite-$N$ as 
well as in the large-$N$ limit. The characteristic feature of the model
is the nonhermiticity induced via the chemical potential, which renders the 
eigenvalues of the Dirac operator complex. Under these conditions one has to 
distinguish in the large-$N$ limit between the regimes of weak and 
strong nonhermiticity \cite{FKS} (see \cite{FS} for a review). 
  
In \cite{A02} a random matrix model with complex eigenvalues  
was proposed for the phase with broken chiral symmetry .
The resulting spectral correlation functions were computed, and
the analytic predictions were confirmed by comparing them with
the results from quenched 
lattice simulations \cite{AWII}, both in the limit of weak and strong 
nonhermiticity. However, the equivalence in the phase with broken symmetry  
between the model proposed in \cite{A02} and the original
model with chemical potential proposed in \cite{Steph} was only 
conjectured \cite{A02,A03}.

In a recent paper \cite{SV03} the authors managed to derive
the spectral density of complex eigenvalues in the regime 
of weak nonhermiticity directly from the effective chiral 
Lagrangian for quenched QCD combined 
with the model \cite{Steph}. Exploiting 
a variant of recently suggested exact replica method
\cite{Kanzieper} they arrived to
a density profile slightly different from earlier 
results of \cite{A02}. Both
results agree asymptotically, and the difference was too small to be 
distinguished from the lattice data \cite{AWII} for the values of chemical  
potential used.  
Under these circumstances it is conceptually  important to be able
to prove the  universality of results within the random matrix model approach,
apart from matching them with first principle lattice data. 
Without such a universality random matrix models loose much of their 
predictive power, being deduced from global symmetry arguments alone.  
When the chemical potential is absent, 
universality was proved in \cite{ADMN,DN}.
For  correlation functions involving complex eigenvalues
only partial results exist for non-chiral random matrix 
models \cite{A02II} at weak nonhermiticity. 
 
One of goals of the present paper is to clarify the issue of equivalence and 
thus possible universality  of the two different random matrix models 
\cite{Steph} and \cite{A02} for QCD with chemical potential, 
at the level of the corresponding partition 
functions. We are going to demonstrate that both models agree at 
the regime of weak nonhermiticity for any 
number of quarks and conjugate anti-quarks, and in that sense 
they are universal.  
At the regime of strong nonhermiticity 
the agreement, however, persists only for an equal number of such  
flavors. For a general flavor content, including only quarks,  
the two partition functions  
\cite{Steph,A02} agree only to the leading order term 
in an expansion in small chemical 
potential. We will also relate our findings to the so far open question 
of universal spectral  eigenvalue correlations, 
by mapping it to the problem of   
universality of the so-called bosonic partition functions.

The object of our investigation can be phrased also in more mathematical 
terms. In the presence of quark flavors the problem of
calculation of the random matrix model 
partition function amounts to computing the expectation value of
a product of characteristic polynomials (also known as spectral determinants) 
in the  model with zero flavors.    
Characteristic polynomials in Hermitian random matrix models have 
received a lot of attention recently, partly due to their relevance 
to the behaviour of Riemann  zeta-function suggested in \cite{KS}. 
On the other hand, such polynomials find important applications
in theory of disordered and chaotic systems, see \cite{AS} and discussion
and further references in \cite{F}.
Different formulas for arbitrary products  
\cite{BH,MN,FW} 
and ratios \cite{F,FS1,SV02,FA,SF2,FS3,BDS}  
of characteristic polynomials have been derived and the universality 
of these expressions has been shown \cite{SF2,AF,V}.  
 
Again much less is known for complex matrix models 
\cite{A01,AVIII,Bergere:2003ht}.  Here one has to distinguish between 
characteristic polynomials and their complex conjugates.  A closed 
determinant formula for arbitrary products of characteristic 
polynomials and their conjugates (of not necessary the same number) 
for quite a general class of models has been given in \cite{AVIII}, see also
related objects emerging in the theory of quantum chaotic scattering
\cite{FK,FS}.  
Our aim here is thus to compute and 
compare such products within the two models 
\cite{Steph,A02}. These results may also be useful when several sets of 
replicas are needed in the computation of two- or higher $k$-point eigenvalue 
correlation functions, generalizing \cite{SV03}.

We derive a compact, new expression for the partition function of
the matrix model \cite{Steph}   at finite-$N$ 
with arbitrary many quark flavours . It
allows us to analyze the chiral phase transition in more detail, 
in particular 
concerning the influence on nondegenerate 
quark masses\footnote{We recall that 
  the model \cite{A02} is always in the broken phase by definition, it 
  cannot reach the phase transition.}.  
We find that the first order phase transition found in \cite{Steph,HJSSV} 
persists, and that it is always driven by the flavor with the smallest (or 
zero) mass. For different mass scales present there is a discontinuity of 
first order for each different mass. This is due to the fact that  
roughly speaking the partition 
function can be written as a determinant over single flavor partition 
functions. A similar phenomenon of having two first order lines for two 
flavors was observed in \cite{KTV}, where two different chemical potentials 
for each flavor were introduced. 
 
The outline of the article is as follows. 
In the section \ref{Zeds} we define the two matrix model partition functions  
for QCD with chemical potential and 
compute them for finite-$N$. We distinguish between the presence of only  
quarks in section \ref{Zq} and quarks with additional complex conjugate  
anti-quarks in section \ref{Zqq}.   
Several technical details of the derivation are summarized in the  
appendices \ref{sigmarep} and \ref{polarJac}.  
In section \ref{Zlarge-N} we turn to the large-$N$ limit, where we distinguish 
between the limit of weak and strong nonhermiticity in sections \ref{weak} and 
\ref{strong}. The resulting consequences for the universality of matrix model 
partition functions are discussed in section \ref{univ}.  
In section \ref{phase} we exploit the results for finite-$N$ from section 
\ref{Zeds} to investigate  
the chiral phase transition at a critical chemical potential, 
for an arbitrary number of quarks.  
In section \ref{disc} we summarize our findings.

\sect{Matrix model partition functions for finite-$N$}\label{Zeds}

\subsection{Partition functions with $N_f$ quark flavors}\label{Zq}

We start with defining the first matrix model for QCD with chemical potential, 
initially introduced by Stephanov \cite{Steph}. 
In the sector of topological charge\footnote{Without loss of generality we 
  restrict ourselves to $\nu\geq0$ throughout the following.}   
$\nu$ it is given by  
\bea 
{\cal Z}^{(N_f,\nu)}_I(\mu;\{m_f\}) \equiv \int d\Phi d\Phi^\dagger 
\prod^{ N_f}_{f=1}  
\det\left( 
\begin{array}{c} 
m_f\one_N                       \ \ \ \     i\Phi + \mu\tilde{\one}_N\\ 
\\ 
i\Phi^\dagger + \mu\tilde{\one}_{N}^\dagger  \ \ \ \    m_f\one_{N+\nu}\\ 
\end{array} 
\right) \exp\left[-N\qq^2\Tr\Phi\Phi^\dagger\right]. 
\label{Z1} 
\eea 
Here $\Phi$ is a complex matrix of size $N\times(N+\nu)$. 
Apart from the unity matrix $\one_n$ of size $n\times n$ we have also  
introduced the rectangular unity matrix of size $N\times (N+\nu)$ 
\be 
(\tilde{\one}_N)_{ij} \ \equiv\  
\left\{ 
\begin{array}{ll} 
\delta_{ij} & i,j=1,\ldots,N \\ 
 0          & j=N+1,\ldots,N+\nu\\ 
\end{array} 
\right. . 
\label{defone} 
\ee 
Eq. (\ref{Z1}) contains $N_f$ quark flavors with real masses $m_f$.  
The chemical potential $\mu$ is added 
to the Dirac matrix of the usual matrix model \cite{SV93} in a standard way
by shifting  
$\Dirac\to\Dirac+\gamma_0\mu$. The model has the  
same global symmetries as QCD with gauge group $SU(N_c\geq3)$ in the  
fundamental representation.  
The Gaussian weight replacing the gauge field average was chosen for  
simplicity in \cite{Steph}. It will allow us to exactly solve the partition 
functions for finite-$N$.  The variance of the random matrix entries 
is such that the Banks-Casher relation for $\mu=0$  is satisfied. 
We will not address the issue of universality by allowing for a more  
general weight function, $\Phi\Phi^\dagger\to V(\Phi\Phi^\dagger)$, with  
$V$ being a polynomial, as in \cite{ADMN,DN}.  
Instead, we compare to a different  
model given in terms of complex eigenvalues \cite{A02} defined below,  
which reduces to the same model \cite{SV93} at $\mu=0$. 
 
In the presence of $\mu\neq0$ a diagonalization necessary to obtain  
complex eigenvalues of the Dirac operator 
does not any longer amounts to a simple procedure.  
More precisely, the angular variables used in the  
singular value decomposition of the matrix $\Phi$ in the form
$\Phi=U_1\Lambda U_2$,with $U_{1,2}$ being unitary, no longer  
decouple in the matrix integral.  Although a Schur decomposition to an 
upper triangular form  
$\phi=U(Z+R)U^\dagger$ remains possible, it does not reveal a natural 
relevant   degrees of freedom in the matrix integral. 
In particular, proceeding in this way one 
retains the  eigenvalues in $Z$ complex
even after setting $\mu=0$, which is not a natural choice of 
integration variables.  
 
The standard fermionic approach to compute the partition function 
remains a viable alternative. It consists in  
replacing the determinants by equivalent Grassmann integrals, 
and further integrating  
out the matrix $\Phi$ explicitly, and finally 
performing a Hubbard-Stratonovich transformation.  
The details are given in the appendix \ref{sigmarep}, 
and include a more general 
case of additional conjugate anti-quarks, to be addressed below.  
For real quark masses the result simplifies considerably  
after choosing a polar decomposition, with the Jacobian computed in   
appendix \ref{polarJac}. After these manipulations
we arrive at the following result 
\bea 
{\cal Z}^{(N_f,\nu)}_I(\mu;\{m_f\}) &\sim& \mbox{e}^{-N\qq^2\Tr\, M^2}  
\int_0^\infty \prod_{f=1}^{N_f}  
dr_k\ r_k^{\nu+1} (r_k^2-\mu^2)^N \mbox{e}^{-N\qq^2 r_k^2} \  
\Delta_{N_f}(r^2)^2 \nn\\ 
&&\times \int dV \int dU \det[U^\dagger]^\nu 
\exp\left[N\qq^2\Tr\left(M U V\hat{r}V^\dagger +   
V\hat{r}V^\dagger U^\dagger M 
\right)\right] . 
\label{Z1UV} 
\eea 
Here, $dU$ and $dV$ denote the Haar measure over the unitary group of  
size $N_f\times N_f$. The matrix $M=\mbox{diag}(m_1,\ldots,m_{N_f})$ 
contains the quark masses and $\hat{r}=\mbox{diag}(r_1,\ldots,r_{N_f})$ is  
the diagonal matrix of radial coordinates.  
$\Delta_{N_f}(r^2)=\prod_{k>l}^{N_f}(r_k^2-r_l^2)$ denotes the  
Vandermonde determinant. 
Due to the unitary invariance we can shift $U\to UV^\dagger$ and then rename  
$U\to U^\dagger$. The resulting unitary integrals can be performed exactly  
using \cite{GW}, together with the fact that $M= M^\dagger$.  
Our first main result valid for finite-$N$ is thus reading   
\bea 
{\cal Z}^{(N_f,\nu)}_I(\mu;\{m_f\}) &\sim&  
\mbox{e}^{-N\qq^2\sum_{f=1}^{N_f}m_f^2}  
\int_0^\infty \prod_{f=1}^{N_f}  
dr_k\ r_k^{\nu+1} (r_k^2-\mu^2)^N \mbox{e}^{-N\qq^2 r_k^2} \ \Delta_{N_f}(r^2) 
\nn\\ 
&&\times \frac{1}{\Delta_{N_f}(m^2)} \  
\det_{i,j=1,\ldots,N_f}\left[I_\nu(2N\qq^2m_i\, r_j)\right]\ . 
\label{Z1ev} 
\eea 
Such a compact expression reducing eq. (\ref{Z1}) to $N_f$ real integrations  
that factorize has previously been known only for $N_f=1$ 
\cite{HJV}. The integral representation for $N_f=2$ in \cite{A03} using 
a Schur decomposition of the matrix $Q=UV\hat{r}V^\dagger$ is more involved. 
Eq. (\ref{Z1ev}) is now amenable to a saddle point computation,  
both at weak and strong nonhermiticity. Furthermore, it will be useful when 
investigating the phase transition in section \ref{phase}. 
 
\indent 
 
We now turn to an alternative random matrix model 
for QCD with chemical potential,  
introduced in \cite{A02} in terms of $N$ complex eigenvalues. 
This type of model can also be 
solved exactly at any finite $N$. 
Here we do not have to distinguish between real 
and complex masses. Moreover, due to the powerful technique of orthogonal  
polynomials the corresponding correlation functions of 
complex eigenvalues also can be found \cite{A02}.  
The model is defined as  
\bea 
{\cal Z}^{(N_f,\nu)}_{II}(\tau;\{m_f\}) &\equiv&  
\int \prod_{j=1}^N 
\left(d^2z_j\ w(z_j,z_j^\ast)\prod_{f=1}^{N_f}m_f^\nu\,(z_j^2+m_f^2)\  
\right) 
\left|\Delta_N(z^2)\right|^2  
\label{Z2}\\ 
w(z,z^\ast) &\equiv& |z|^{2\nu+1}\! \exp\left[ 
-\frac{N}{1-\tau^2}\left(|z|^2 -\frac{\tau}{2}(z^2+z^{2\,\ast})\right) 
\right]\ . 
\label{weight} 
\eea 
The parameter $\tau\in[0,1]$ appearing in the Gaussian weight function  
$w(z,z^\ast)$ controls the effective degree of nonhermiticity. It allows to   
interpolate between models with real and maximally 
complex eigenvalues for $\tau=1$ and 
$\tau=0$, respectively. The two partition functions eq. (\ref{Z1}) and eq.  
(\ref{Z2}) are of course different in general at finite $N$. Only in the  
Hermitian limit 
$\tau \to 1$, $\mu \to 0$
they are exactly the same   
under the identification of the masses: $m_{f,II}=m_{f,I} \qq \sqrt{2}$.  
In fact, in this paper we will show a much more interesting relation:  two
models are also generally equivalent at an appropriate large $N$ limit,
under an appropriate   correspondence among the relevant parameters 
(the mapping    between the two models actually depends on 
the way the large $N$ limit is performed).  
 
The massive partition function eq. (\ref{Z2})  
can be evaluated due to the following observation.  
We can write  
\be 
{\cal Z}^{(N_f,\nu)}_{II}(\tau;\{m_f\})\ \sim\  
\left\langle  \prod_{j=1}^N \prod_{f=1}^{N_f}m_f^\nu \, 
(z_j^2+m_f^2)\right\rangle \ , 
\label{Z2vev} 
\ee 
where the expectation value is taken with respect to the zero flavor partition 
function, ${\cal Z}^{(N_f=0,\nu)}_{II}(\tau)$. 
The relation to a product of characteristic polynomials is evident. 

The relevant set of orthogonal polynomials with 
respect to the weight eq. (\ref{weight}) 
are given by the complex generalization of standard
Laguerre polynomials, see \cite{A02}  :
\be 
\tilde{P}_k^{(\nu)}(z^2)\ \equiv\  
(-1)^k k!\left(\frac{2\tau}{N}\right)^k  
L_k^{(\nu)}\left(\frac{Nz^2}{2\tau}\right) \ ,  
\label{Laguerre} 
\ee 
which are given here in the monic normalization.  
Using the theorem proven in \cite{AVIII} the expectation value  
eq. (\ref{Z2vev}) can be conveniently expressed as 
a determinant of size $N_f\times N_f$ , with entries being
these orthogonal polynomials.  
Taking into account the extra factors 
$\prod_{f=1}^{N_f}m_f^\nu$ we arrive at \cite{AVIII} 
\be 
{\cal Z}^{(N_f,\nu)}_{II}(\tau;\{m_f\})\  
\sim\ \frac{\det_{k,l=1,\ldots,N_f}\left[ 
m_l^\nu\tilde{P}_{N+k-1}^{(\nu)}(-m_l^2) 
\right]}{\Delta_{N_f}(m^2)}\ . 
\label{Z2N} 
\ee 
We would like to note that the 
partition function is real for real quark masses  
as it should be. This is not 
obvious at all when looking at the definition eq. (\ref{Z2}).  
The obtained result is exact for finite-$N$. 
It can also be continued to complex quark 
masses without modification.

\subsection{Partition functions with quarks and complex conjugate anti- 
quarks}\label{Zqq}

In this subsection we enlarge the flavor space by adding pairs of 
complex conjugate 
quarks to the partition functions eqs. (\ref{Z1}) and (\ref{Z2}).  
Such partition functions were already considered for one such pair in 
\cite{Steph} when analyzing the quenched approximation and for $n$ such pairs 
of degenerate mass in \cite{SV03} in the replica approach. 
The special feature 
of quarks and conjugate anti-quarks occurring in the partition function 
at the same time is that   they may form 
a non vanishing meson density. This is, in turn,
reflected in the existence of an 
effective chiral Lagrangian with isospin chemical potential 
as pointed out in \cite{TV}, linking it to QCD. 
 
Here, we will derive such an effective model 
in terms of a unitary group integral 
over the Goldstone manifold directly from the underlying
random matrix model.  
We will allow for any, not necessarily equal  
number of quarks and conjugate anti-quarks 
with non degenerate mass.  
In the first model 
eq. (\ref{Z1}) the presence of additional conjugate 
anti-quarks implies considerably  
more effort  in computing the partition functions. In the procedure we
will make use of the results  from \cite{SV03}. 
In contrast to that in the second model eq. (\ref{Z2})   
the resulting partition functions immediately 
follow from the theorem proved  in \cite{AVIII}. 
 
The matrix model partition functions eq. (\ref{Z1}) with $m$ quarks of mass 
$m_f$ and $n$ conjugate anti-quarks of mass $n_f^*$ in the sector of  
topological 
charge $\nu$ is defined as  
\bea 
&&{\cal Z}^{(N_f=m+n,\nu)}_I(\mu;\{m_f\}_m,\{n_g^*\}_n)  
\ \equiv  
\label{Z1qq}\\  
&&\equiv\int d\Phi d\Phi^\dagger  
\prod^{ m}_{f=1}  
\det\left(\! 
\begin{array}{c} 
m_f\one_N                       \ \ \ \     i\Phi + \mu\tilde{\one}_N\\ 
\\ 
i\Phi^\dagger + \mu\tilde{\one}_{N}^\dagger  \ \ \ \    m_f\one_{N+\nu}\\ 
\end{array} 
\!\right)  
\prod^{ n}_{g=1}  
\det\left(\! 
\begin{array}{c} 
n_g^*\one_N                       \ \ \ \     -i\Phi + \mu\tilde{\one}_N\\ 
\\ 
-i\Phi^\dagger + \mu\tilde{\one}_{N}^\dagger  \ \ \ \    n^*_g\one_{N+\nu}\\ 
\end{array} 
\!\right)  
\mbox{e}^{-N\qq^2\Tr\Phi\Phi^\dagger}\!\! . 
\nn 
\eea 
Note that in contrast to the previous section we now allow 
for complex masses, as they may serve as source terms for the complex 
Dirac operator eigenvalues (see e.g. in \cite{SV03}). 
Using the standard fermionization technique we arrive 
at the following matrix model representation 
in terms of the complex matrix  $Q$ of size $N_f\times N_f$ 
\bea 
{\cal Z}^{(N_f,\nu)}_I(\mu;M)&\sim& 
\int dQ dQ^\dagger  
\det[M+Q^\dagger]^\nu  
\label{Z1qqQm}\\ 
&&\times\det\left[(M+Q^\dagger)(M+Q) 
-\mu^2(M+Q^\dagger)\Sigma_3(M+Q^\dagger)^{-1}\Sigma_3\right]^N 
\mbox{e}^{-N\qq^2\Tr \, Q^\dagger Q}, 
\nn 
\eea 
with the mass matrix $M=$ diag$(m_1,\ldots,m_m,-n_1^*,\ldots,-n_n^*)$ 
and $\Sigma_3=$diag$(\one_m,-\one_n)$ being a generalized Pauli matrix.  
The details of the derivation are given in appendix \ref{sigmarep}. It 
is interesting here to work out the limit of real masses, 
$M=M^\dagger$ explicitly.  
To this end, we introduce a polar decomposition of the matrix 
$Q=U\,R$ with $R=V\hat{r}V^\dagger$ being hermitian with positive 
eigenvalues $r_k$, and $U,V$ being unitary. Using eq. (\ref{Z1Q}) and 
the Jacobian calculated in the appendix \ref{polarJac} we obtain 
\bea 
{\cal Z}^{(N_f,\nu)}_I(\mu;M)&\sim& \mbox{e}^{-N\qq^2\Tr\, M^2}  
\int_0^\infty \prod_{f=1}^{N_f}  
dr_k\ r_k^{\nu+1} \mbox{e}^{-N\qq^2r_k^2} \  
\Delta_{N_f}(r^2)^2 \int dV \int dU \det[U^\dagger]^\nu 
\nn\\ 
&\times&  
\mbox{e}^{N\qq^2\Tr M (U V\hat{r}V^\dagger + 
V\hat{r}V^\dagger U^\dagger)} 
\det\left[V\hat{r}^2V^\dagger 
-\mu^2 V\hat{r}V^\dagger U^\dagger\Sigma_3 UV\hat{r}^{-1}V^\dagger\Sigma_3  
\right]^N . 
\nn\\ 
\label{Z1qqUR} 
\eea 
This expression will be used in the next section.

\indent 
 
Let us now consider the second matrix model \cite{A02} with
pairs of complex conjugate 
quarks inserted. The corresponding partition function is defined as  
\bea 
{\cal Z}^{(N_f,\nu)}_{II}(\tau;\{m_f\}_m,\{n_g^*\}_n) &\equiv& \! 
\int \prod_{j=1}^N 
d^2z_j\ w(z_j,z_j^\ast)\prod_{f=1}^{m}m_f^\nu(z_j^2+m_f^2) 
\prod_{g=1}^{n}n_g^{*\,\nu}(z_j^{*\,2}+n_g^{*\,2}) 
\left|\Delta_N(z^2)\right|^2  
\nn\\ 
&\sim&  
\left\langle  \prod_{j=1}^N \left(\prod_{f=1}^{N_f}m_f^\nu  
(z_j^2+m_f^2)\prod_{g=1}^{n}n_g^{*\,\nu}(z_j^{*\,2}+n_g^{*\,2}) 
\right)\right\rangle \ , 
\label{Z2qq} 
\eea 
where we again presented this object in a form of an 
expectation value.  
In contrast to eq. (\ref{Z2qq}) we do not need to distinguish 
between signs of 
masses and chemical potential for the quarks and their conjugates.  
The reason is that the large-$N$  
result for eq. (\ref{Z2qq}) turns out to be a quadratic 
function in all masses. Apart from that, in  
the identification between the nonhermiticity 
parameters $\tau$ and $\mu$ appears quadratically.  
The same weight function eq. (\ref{weight}) for all eigenvalues is therefore 
equally valid for both signs of the chemical potential\footnote{In 
this sense the model does not 
  distinguish between isospin and baryon chemical potential.}. 
 
The evaluation of eq. (\ref{Z2qq}) is again straightforward, due to the 
general theorem proved in \cite{AVIII}.  
To write the result in a compact form we need to introduce more notation.  
In addition to the orthogonal polynomials eq. (\ref{Laguerre}) 
the so-called  bare kernel made of these polynomials appears 
in the corresponding expressions:
\bea 
\kappa_N(z^2,u^{*\,2}) &\equiv&  
\sum_{k=0}^{N-1}P_k^{(\nu)}(z^2)P_k^{(\nu)}(u^{*\,2})\nn\\ 
&=&  
\frac{1}{f^{(\nu)}(\tau)} 
\sum_{k=0}^{N-1} \frac{\Gamma(\nu+1)\,k!}{\Gamma(\nu+k+1)}\tau^{2k} 
L_k^{(\nu)}\left(\frac{Nz^2}{2\tau}\right) 
L_k^{(\nu)}\left(\frac{Nu^{*\, 2}}{2\tau}\right)\ . 
\label{kernel} 
\eea 
The kernel contains a sum over the orthonormalized polynomials 
$P_k^{(\nu)}(z^2)$,  
\be 
P_k^{(\nu)}(z^2) \ \equiv\ h_k^{-\frac12} \tilde{P}_k^{(\nu)}(z^2)\ , 
\label{OP} 
\ee 
with norms given by
\be 
h_k \ \equiv\ \int d^2 z \ w(z,z^*)\tilde{P}_k^{(\nu)}(z^2) 
\tilde{P}_k^{(\nu)}(z^{*\,2}) 
\ =\  
f^{(\nu)}(\tau) \left( \frac{2}{N}\right)^{2k}  
 \frac{\Gamma(\nu+1+k)\,k!}{\Gamma(\nu+1)}\ . 
\label{norm} 
\ee 
Here we 
have introduced the functions
\be 
f^{(\nu)}(\tau)\ \equiv\ \int d^2 z \ w(z,z^*)\ =\  
\pi\Gamma\left(\nu+\frac32\right)(1-\tau^2)^{\frac{\nu}{2}+\frac34} \  
{\cal  P}_{\nu+\frac12}\left( \frac{1}{\sqrt{1-\tau^2}}\right)\ , 
\label{f} 
\ee 
where ${\cal  P}_\gamma (x)$ stands for the Legendre function. 
The full kernel $K_N(z,u^*)$ as it appears in the expression for the 
eigenvalue correlation functions is then obtained by multiplying 
the bare kernel with the 
weight functions, 
$K_n(z,u^*)\equiv[w(z,z^*)w(u,u^*)]^{\frac12}\kappa_N(z^2,u^{*\,2})$.  
Following \cite{AVIII} we immediately obtain  
\bea 
{\cal Z}^{(N_f,\nu)}_{II}(\tau;\{m_f\}_m,\{n_g^*\}_n) &\sim&  
\frac{\prod_{f=1}^m m_f^\nu\ \prod_{g=1}^n n_g^{*\,\nu} 
}{\Delta_m(m^2)\Delta_n(n^{*\, 2})} 
\left(\prod_{k=N}^{N+m-1}\!h_k\right) 
\det_{i,j=1,\ldots,n}[{\cal   B}(m_i,n_j^*)] \ ,  
\label{Z2qqN}\\ 
{\cal  B}(m_i,n_j^*) &\equiv& \left\{ 
\begin{array}{ll} 
\kappa_{N+m}(-m_i^2,-n_j^{*\,2})              & i=1,\dots,m \\ 
\tilde{P}_{N+i-1}^{(\nu)}(-n_j^{*\,2})     & i=m+1,\ldots,n 
\end{array} 
\right. \ . 
\label{B} 
\eea 
where we may assume that $n\geq m$, 
without loss of generality (the case $n<m$ 
follows from complex conjugation).   
This result is exact for finite-$N$.
We note that the model eq. (\ref{Z2qq}) (and 
eq. (\ref{Z2})) is always in the phase with broken chiral symmetry \cite{A02}.

\sect{The large-$N$ limit}\label{Zlarge-N} 
 
\subsection{The weak nonhermiticity limit}\label{weak} 
 
The limit of weak nonhermiticity \cite{FKS} is defined by  
taking simultaneously  
the large-$N$ limit $N\to\infty$ and the hermitian limit $\tau\to1$ or  
$\mu\to0$ in such a way that the following product is kept constant: 
\bea 
\lim_{N\to\infty,\, \tau\to1} 
N(1-\tau^2)&\equiv& \alpha^2 
\eea 
or, equivalently
\bea
\lim_{N\to\infty,\, \mu\to0}2N\qq^2\mu^2 &=&   \tilde{\alpha}^2\ .  
\label{weaklim} 
\eea 
In this limit the macroscopic spectral density has support only on the real  
line and is given by a semicircle for both models eqs. (\ref{Z1}) and  
(\ref{Z2}), in the limit $\tau\to1$ and $\mu\to0$ respectively \cite{A02}.  
In contrast  
to that the microscopic correlation functions differ from those on the real 
line. They still extend into the complex plane and depend explicitly 
on the parameter $\al$ or $\tilde{\al}$. To identify them for the two models  
a relation between $\tau$ and $\mu$, or $\al$ and $\tilde{\al}$, has to be 
imposed, which we will find by comparing the two partition functions.

We will see that also the partition functions at weak nonhermiticity contain 
important information, and this has already been 
exploited for example in \cite{SV03}   when computing
the microscopic density.  
It turns out that the partition functions 
with quarks alone differ from the partition functions at  
$\mu=0$ only by
an overall prefactor $\exp[-N_f\al/2]$.
When adding conjugate anti-quarks the situations however
changes and the partition
functions differ from their $\mu=0$ values 
nontrivially\footnote{Obviously, for only conjugate  
anti-quarks alone we are back to the situation of only quarks.}.

The microscopic rescaling of the quark masses (and complex eigenvalues)  
at weak nonhermiticity is defined as 
\bea 
\zeta_f &\equiv& 2N\qq m_f \ , f=1,\ldots,m \nn\\ 
\xi_g^* &\equiv& 2N\qq n_g^*\ , g=1,\ldots,n 
\label{microweak} 
\eea 
where $2N\qq\sim\rho(0)$ is the macroscopic density at the origin of  
the model eq. (\ref{Z1}), after taking the limit $\mu\to0$.  
 
We begin with the partition functions containing only quarks from subsection 
\ref{Zq}.  First of all due to the rescaling of the masses eq.  
(\ref{microweak}) the constant prefactor $\mbox{e}^{-N\qq^2\Tr\, M^2}$ 
reduces to unity. 
Taking the weak limit eq. (\ref{weaklim}) we can replace the factor 
\be 
(r_k^2-\mu^2)^N \ \to\ r_k^{2N} 
\exp\left[-N\mu^2r_k^{-2}\right] 
\label{SPlim} 
\ee 
inside the integral eq. (\ref{Z1ev}) over the radial coordinates.  
After rescaling the arguments of the  
Bessel functions through eq. (\ref{microweak}) 
a saddle point evaluation in the variables $r_k$ 
leads to the value 
\be 
\hat{r}\arrowvert_{\mbox{sp}}\ =\ \one_{N_f} \qq^{-1}\ . 
\label{weakSP} 
\ee 
Since the determinant in eq. (\ref{Z1ev}) is degenerate at the saddle point we 
have to take into account the Gaussian fluctuations. Performing the
calculation, we get
\be 
{\cal Z}^{(N_f,\nu)}_I(\mu;\{\zeta_f\})\arrowvert_{\mbox{weak}}\ \sim\   
\exp\left[-N_f\qq^2 N\mu^2\right]\ 
\frac{\det_{k,l=1,\ldots,N_f}\left[\zeta^{k-1}_l I_{\nu}^{(k-1)}(\zeta_{l}) 
\right]}{\Delta_{N_f}(\zeta^2)}\ . 
\label{Z1weak} 
\ee 
With $I_\nu^{(j)}(x)$ we denote the $j-$th derivative of the modified Bessel 
functions.   
This final expression for the partition function generalizes previous results 
which were known only in the sector of topological charge $\nu=0$ with   
$N_f=1$ \cite{HJV} and $N_f=2,3$ flavors \cite{A03}.  
Eq. (\ref{Z1weak}) 
is exactly the same as for $\mu=0$ \cite{JSV,BRT}, apart from the  
exponential prefactor $\exp[-N_f\tilde{\al}^2/2]$.  
This implies in particular that the partition functions 
obey the same consistency conditions \cite{AD} as those for $\mu=0$. 
More generally speaking, they belong to the same Toda lattice hierarchy as it 
was already exploited in \cite{SV03}.

\indent 
 
We turn to the second partition function eq. (\ref{Z2N}).  
In order to read off the proper rescaling defined in eq. (\ref{microweak})  
we first take the {\it Hermitian}  
limit $\tau\to1$ of the weight eq. (\ref{weight}) 
in order to determine the corresponding variance. 
This procedure is based on the macroscopic spectral
density at weak nonhermiticity being given by taking the Hermitian 
limit\footnote{The corresponding weak nonhermiticity limit 
for the weight function is given in section 
\ref{univ} eq. (\ref{weakweight}), but we do not need it  
here.}:  
\bea 
\lim_{\tau\to1} w(z,z^\ast) &=&  
\lim_{\tau\to1}\  (x^2+y^2)^{\nu+\frac12}\exp\left[-\frac{N}{1+\tau}x^2 
-\frac{N}{1-\tau}y^2\right] 
\nn\\ 
&=& x^{2\nu+1}\ \sqrt{\frac{\pi\al^2}{2N^2}}\  
\delta\left(y\right)  
\exp\left[-\frac12N x^2\right]\ , 
\label{hermweight} 
\eea 
where we have inserted $z=x+iy$. Therefore we have to use the value  
$\qq=1/\sqrt{2}$ in eq. (\ref{microweak}), leading to the correspondence
$\zeta_f=\sqrt{2}Nm_f$ and similar for the conjugate. Such a rescaling  
which is different for the two models is precisely the mapping  
between the two sets of different mass parameters mentioned after  
eq. (\ref{weight}). 
 
We could have also introduced a  
variable $\qq$ in the model  eq. (\ref{Z2N}) by rescaling the eigenvalues.  
However, since the parameter drops out after microscopic rescaling  
we kept $\qq^2=1/2$ for simplicity here.  
In order to perform the weak nonhermiticity limit in eq. (\ref{Z2N}) we first 
extract the powers of $\tau$ used 
in the definition of the polynomials (\ref{Laguerre}) 
from the corresponding determinant.  
Using the definition $\tau^2=1-\al^2/N$ we obtain the prefactor 
\be 
\lim_{N\to\infty\,\tau\to1} \tau^{N_fN+N_f(N_f-1)/2}\ =\  
\exp\left[-\frac12 N_f\al^2\right]\ . 
\label{weaktau} 
\ee 
The remaining determinant of Laguerre polynomials can be evaluated 
in the same way as for 
$\mu=0$ using eq. (\ref{microweak}) 
\be 
\lim_{N\to\infty,\,\tau\to1} m^\nu L_N^{(\nu)}\left(\frac{-Nm^2}{2\tau}\right) 
\ \sim\ I_\nu(\zeta)\ , 
\label{Laguerreasymp} 
\ee 
and differentiating due to degeneracy. Ignoring constant factors and powers 
of $N$ we thus obtain 
\be 
{\cal Z}^{(N_f,\nu)}_{II}(\al;\{\zeta_f\})\arrowvert_{\mbox{weak}}\ \sim\  
\exp\left[-\frac12 N_f\al^2\right]\  
\frac{\det_{k,l=1,\ldots,N_f}\left[\zeta^{k-1}_l I_{\nu}^{(k-1)}(\zeta_{l}) 
\right]}{\Delta_{N_f}(\zeta^2)}\ . 
\label{Z2weak} 
\ee 
We immediately see that the two partition functions (\ref{Z1weak}) 
and (\ref{Z2weak}) agree upon identifying 
\be 
(1-\tau^2)\ =\ 2\qq^2\mu^2 
\label{mutau1} 
\ee 
or, equivalently $\al=\tilde{\al}$. 
Such an identification can be conveniently interpreted in terms of
equating the two macroscopic densities in the complex plane 
for small $\mu$  \cite{A02}. In principle,
partition functions can always be multiplied by an 
overall constant, making them agree. 
The nontrivial statement is that in both 
cases the rescaled nonhermiticity parameter factors out.

We thus proved the equivalence of the two partition functions  
is the limit of weak nonhermiticity, for an arbitrary number of flavors  
$N_f$ of nondegenerate masses $m_f$ for any given topological sector $\nu$. 
This equivalence was claimed in \cite{A03} and verified there for degenerate 
masses and sector $\nu=0$, up to and including $N_f=3$ flavors.

\indent

Next we investigate the weak nonhermiticity limit of 
the partition function with 
quarks and conjugate anti-quarks, eq. (\ref{Z1qqQm}). In this case, all 
powers higher than linear in $M$ and $\mu^2$ in eq. (\ref{Z1qqQm}) are 
dropped. After expanding also the determinant to the power $N$ in powers of 
$M$, we have  
\bea 
{\cal Z}^{(N_f,\nu)}_I(\mu;M)&\sim& 
\int dQ dQ^\dagger  
\det[Q^\dagger]^\nu \mbox{e}^{-N\qq^2\Tr \, Q^\dagger Q}  
\label{Z1qqQexp}\\ 
&&\times \exp\left[N\Tr\ln(Q^\dagger Q)+N\Tr\left(M(Q^{-1}+Q^{\dagger\,-1}) 
-\mu^2 Q^{-1}\Sigma_3 Q^{\dagger\,-1}\Sigma_3\right) + {\cal O}(1/N)\right] 
\!\! . 
\nn 
\eea 
By using again the polar decomposition $Q=U\,R$ and the respective  
Jacobian, we obtain  
\bea 
{\cal Z}^{(N_f,\nu)}_I(\mu;M)&\sim&\! 
\int_0^\infty \prod_{f=1}^{N_f}  
dr_k\ r_k^{\nu+1} \mbox{e}^{-N\qq^2 r_k^2+N\ln r_k^2} \  
\Delta_{N_f}(r^2)^2 \int dV \int dU \det[U^\dagger]^\nu 
\label{Z1qqUV}\\ 
&\times& \! 
\exp\left[N\Tr\left(M (V\hat{r}^{-1}V^\dagger U^\dagger + 
UV\hat{r}^{-1}V^\dagger) 
\right)  
-\mu^2 N\Tr\left(V\hat{r}^{-1}V^\dagger U^\dagger\Sigma_3 
UV\hat{r}^{-1}V^\dagger\Sigma_3 
\right)\right] . 
\nn 
\eea 
In this form the integral is amenable to a saddle point approximation at 
large-$N$ in the variables $r_k$, which will lead to the chiral Lagrangian 
picture \cite{TV}.  
In the previous case, with only quarks included 
the $\mu$-dependence factorizes out as we discussed above. 
Now the dependence will be less trivial, making it a 
real check for the equivalence of the two models. 
Due to the rescaling eqs. (\ref{microweak}) and (\ref{weaklim}) the exponents 
inside the unitary integral are of the order of unity and the the 
saddle point value
is taken at 
\be 
\hat{r}\arrowvert_{\mbox{sp}}\ =\ \one_{N_f} \qq^{-1}\ , 
\ee 
as previously in eq. (\ref{weakSP}). Since the unitary integral is non 
vanishing at $\hat{r}\arrowvert_{\mbox{sp}}\sim\one_{N_f}$, we have 
\bea 
{\cal Z}^{(N_f,\nu)}_I(\mu;M)\arrowvert_{\mbox{sp}} &\sim& 
\int dU \det[U^\dagger]^\nu 
\exp\left[N\qq\Tr(M (U^\dagger + U))  
-\mu^2 N\qq^2\Tr(U^\dagger\Sigma_3 U\Sigma_3)\right] .\nn\\ 
&& 
\label{Z1qqUVSP} 
\eea 
This is precisely (a zero-dimensional version of) 
the chiral effective theory for isospin chemical potential 
as derived in \cite{TV}, with replacements $2N\longleftrightarrow V$ and  
$\mu^2\qq^2\longleftrightarrow \mu^2F_\pi^2$.  
The manifestation of the zero-dimensional nature is that the
parameter $F_\pi$,   the pion decay constant entering in the
chiral Lagrangian \cite{TV}, is not contained in the present matrix 
model partition function, as it comes from the non-zero momentum modes of the 
Goldstone bosons.  
If we wish to write the $\mu-$dependent term as  
$\frac12\Tr[U^\dagger,\Sigma_3] [U,\Sigma_3]$ as it appears when deriving it 
from a gauge principle, we have to multiply a factor  
$\exp\left[N\qq^2\mu^2N_f\right]$ to the partition function. By doing so also 
in the previous section  
it would remove the trivial $\mu-$dependence, that  
leads to an unphysical negative quark 
number density in the broken phase  
(see e.g. in \cite{HJV} for $N_f=1$).  
Finally we wish to mention that the usual mass term in the unquenched,  
effective QCD Lagrangian is $\Tr(M^\dagger U^\dagger +MU)$ instead. There  
complex masses are needed to locate the dependence of the $\theta-$angle in 
the action for example, leaving the partition function real.

We are now ready to evaluate the unitary integral in eq. (\ref{Z1qqUVSP}) 
exactly, following closely \cite{SV03}. There, the following  
parameterization of 
the unitary matrix $U\in U(N_f=m+n)$ has been suggested\footnote{In addition 
  to \cite{SV03} we have performed an extra rotation $u_i\to v_i u_i 
  v_i^\dagger$ for $i=1,2$, interchanging the first 2 matrices.} 
\bea 
U &=&  
\left( 
\begin{array}{cc} 
v_1 & 0\\ 
0   & v_2\\ 
\end{array} 
\right) 
\left( 
\begin{array}{cc} 
u_1 & 0\\ 
0   & u_2\\ 
\end{array} 
\right) 
\Lambda 
\left( 
\begin{array}{cc} 
v_1^\dagger &           0\\ 
0           & v_2^\dagger\\ 
\end{array} 
\right) \ , 
\label{Upar}\\ 
\Lambda&\equiv& 
\left( 
\begin{array}{ccc} 
\hat{\la}            & \sqrt{\one_m-\hat{\la}^2} & 0\\ 
\sqrt{\one_m-\hat{\la}^2} & -\hat{\la}            & 0\\ 
0                    & 0                    &-\one_{n-m}\\ 
\end{array} 
\right)\ . 
\label{Lapar} 
\eea 
Here, we denote 
$\hat{\la}\equiv$ diag$(\la_1,\ldots,\la_m)$ with $\la_k\in[0,1]$ for 
$k=1,\ldots,m$. The unitary submatrices are $u_1,v_1\in U(m)$, $u_2\in U(n)$ 
and $v_2\in \tilde{U}(n)\equiv U(n)/(U(1)^m\times U(n-m))$,  
where $n\geq m$ has been chosen 
without loss of generality. The Jacobian for this transformation was computed 
in \cite{SV03}, 
\be 
J(\{\la_k\})\ \equiv\ \prod_{k>l}^m(\la_k^2-\la_l^2)^2\  
\prod_{j=1}^m 2\la_k(1-\la_k^2)^{n-m}\ . 
\label{laJac} 
\ee 
For the trace containing $\Sigma_3$ it follows 
\be 
\Tr(U^\dagger\Sigma_3 U\Sigma_3) \ =\  
n-3m +4\sum_{k=1}^m \la_k^2 \ . 
\label{TrSig} 
\ee 
With these steps taken we can insert the above parameterization  
in eq. (\ref{Z1qqUVSP}),  
arriving at the following factorized group integrals  
\bea 
{\cal Z}^{(N_f,\nu)}_I(\mu;\{m_f\}_m,\{n_g^*\}_n)&\!\!\!\! 
\arrowvert_{\mbox{weak}}& \ \sim 
\exp\left[-(n-3m)N\qq^2\mu^2\right]\int_0^1 \prod_{k=1}^m d\la_k J(\{\la_k\}) 
\mbox{e}^{-4\qq^2N\mu^2\la_k^2} 
\nn\\ 
&\!\!\!\!\times&\!\!\!\!\!\!\int_{U(m)}\!du_1\int_{U(m)}\!dv_1 \det[u_1]^\nu 
\exp\left[N\qq\Tr(u_1^\dagger v_1^\dagger \hat{m}\,v_1\hat{\la}+ v_1^\dagger 
\hat{m}\,v_1u_1\hat{\la})\right]  
\nn\\ 
&\!\!\!\!\times&\!\!\!\!\!\!\int_{U(n)}\!du_2\int_{\tilde{U}(n)} 
\!dv_2 \det[u_2]^\nu 
\exp\left[N\qq\Tr(u_2^\dagger v_2^\dagger \hat{n}^*v_2\hat{\la}_-+ 
v_2^\dagger \hat{n}^*v_2u_2\hat{\la}_-)\right]\! . 
\nn\\ 
\label{Z1qqint} 
\eea 
By splitting into blocks we introduced the following obvious notation,  
denoting  
$\hat{m}\equiv$ diag$(m_1\ldots,m_m)$, 
$\hat{n}^*\equiv$ diag$(-n_1^*\ldots,-n_n^*)$, and by   
$\hat{\la}_-\equiv$ diag$(-\la_1,\ldots,-\la_m,-1,\ldots,-1)$ 
a matrix of size $n\times n$.  
Performing the additional transformations $u_i\to v_i^\dagger u_i$  
and renaming $u_i\to u_i^\dagger$ for $i=1,2$, we can bring both double 
unitary integrals to the form 
\bea 
&&\int_{U}du\int_{U}dv \det[u\,v]^\nu 
\exp\left[\frac12\Tr(u^\dagger\hat{a}\,v^\dagger\hat{b}+ v 
\hat{a}\,u\hat{b})\right] \ =  
\frac{1}{\Delta(a^2)\Delta(b^2)}\det_{i,j}[I_\nu(a_ib_j)] \ , 
\label{BTint} 
\eea 
with $U=U(l)$ for $l=m$ or $n$, respectively\footnote{In order to apply the 
  result \cite{SW}  
  we have to promote $v_2$ to the full unitary group $U(n)$. We first multiply 
  by the additional integrations 
  $(\int_{U(1)}dw_i\det(w_i)^\nu)^m\int_{U(n-m)}dw_0\det(w_0)^\nu$ and shift  
$u_2\to$diag$(w_1,\ldots,w_m,w_0)u_2$. Due to the cyclicity of the trace and 
  the $(n-m)-$fold degeneracy of $\hat{\lambda}_-$ which makes it commute with 
 diag$(w_1,\ldots,w_m,w_0)$ we obtain a matrix $v_2$diag$(w_1,\ldots,w_m,w_0)$ 
that parameterizes the full group $U(n)$.}. 
Here we have used the result \cite{SW} which gives this integral also in the 
case when the matrices $\hat{a}$ and $\hat{b}$ are not hermitian. In our 
special case the $a_i^2$ and $b_i^2$ are the complex eigenvalues of the 
diagonal matrices  $\hat{a}^2$ and $\hat{b}^2$ respectively.  
In the second double unitary group integral we still have to take limits due 
to the $(n-m)$-fold degeneracy in the matrix $\hat{\la}_-$. Both integrals 
together then cancel the Jacobian $J(\{\la_k\})$ up to the factor  
$\prod_{k=0}^m 2\la_k$, leading to  
\bea 
&&{\cal Z}^{(N_f,\nu)}_I(\mu;\{\zeta_f\}_m,\{\xi_g^*\}_n) 
\arrowvert_{\mbox{weak}}\  
\sim\  
\frac{\exp[-(n-3m)\qq^2N\mu^2] }{\Delta_m(\zeta^2)\Delta_n(\xi^{*\,2})}  
\int_0^1 \prod_{k=1}^m d\la_k\la_k \mbox{e}^{-4\qq^2N\mu^2\la_k^2} 
\nn\\ 
&&\times\det_{k,l=1,\ldots,m}[I_\nu(\la_k\zeta_l)]\   
\det 
\left( 
\begin{array}{cccccc} 
I_\nu(\xi_1^*\la_1) &\cdots& I_\nu(\xi_1^*\la_m) &I_\nu(\xi_1^*)& \cdots&  
\xi_1^{*\,n-m-1}I_\nu^{(n-m-1)}(\xi_1^*)\\ 
\vdots              &      &  \vdots             & \vdots       &       & 
\vdots \\   
I_\nu(\xi_n^*\la_1) &\cdots& I_\nu(\xi_n^*\la_m) &I_\nu(\xi_n^*)& \cdots&  
\xi_n^{*\,n-m-1}I_\nu^{(n-m-1)}(\xi_n^*)\\ 
\end{array} 
\!\right). 
\nn\\ 
&& 
\label{Z1qqdet^2} 
\eea 
The above equation can be further simplified. First we multiply the 
two determinants of Bessel functions,  
$\det(B)\det(A)=\det(AB)$. Because of the different size of 
the two matrices we have to add a block of unity $\one_{n-m}$ to the first 
matrix $I_\nu(\la_k\zeta_l)$ of size $m\times m$. Each entry  
in the first $m$ columns of the resulting product matrix $AB$ contains a sum 
of $m$ terms, $\sum_{j=1}^m I_\nu(\xi^*_k\la_j)I_\nu(\la_j\zeta_l)$. Due to 
the symmetry of the integrand under permutations of the variables $\la_k$ and 
due to the invariance properties of determinants the sums can be reduced to 
single terms, $I_\nu(\xi^*_k\la_k)I_\nu(\la_k\zeta_l)$, and we only sketch 
the procedure briefly.  
We begin with the first column. As determinants differing only by one column 
can be added, we expand  
the determinant as a sum of $m$ terms, which only differ from each other
by the labeling of $\la_k$ in the first column. Due to the invariance 
with respect to permuting $\la_k$ 
all terms can be written as $m$ times 
the same determinant, with label $\la_1$ in the 
first column. Next we can eliminate all the terms with label $\la_1$ in the 
remaining columns, by successively subtracting the first column times an 
appropriate factor. Next we process the second column in the same way, keeping 
only the label $\la_2$ there, and so forth.  
 
We finally arrive at the following 
expression 
\bea 
&&{\cal Z}^{(N_f=m+n,\nu)}_I(\mu;\{\zeta_f\}_m,\{\xi_g^*\}_n) 
\arrowvert_{\mbox{weak}} \ \sim 
\frac{\exp\left[-(n-3m)\qq^2N\mu^2\right]}{ 
\Delta_m(\zeta^2)\Delta_n(\xi^{*\,2})} \times 
\nn\\ 
&& 
\det 
\left(\!\!\! 
\begin{array}{ccc} 
\int_0^1d\la_1\la_1 
\mbox{e}^{-4N\qq^2\mu^2\la_1^2}I_\nu(\xi_1^*\la_1)I_\nu(\la_1\zeta_1)  
&\cdots& 
\int_0^1d\la_1\la_1 
\mbox{e}^{-4N\qq^2\mu^2\la_1^2}I_\nu(\xi_n^*\la_1)I_\nu(\la_1\zeta_1)\\  
\vdots              &      &  \vdots \\   
\int_0^1d\la_m\la_m 
\mbox{e}^{-4N\qq^2\mu^2\la_m^2}I_\nu(\xi_1^*\la_m)I_\nu(\la_m\zeta_m)  
& \cdots& 
\int_0^1d\la_m\la_m 
\mbox{e}^{-4N\qq^2\mu^2\la_m^2}I_\nu(\xi_n^*\la_m)I_\nu(\la_m\zeta_m)\\  
I_\nu(\xi_1^*) &\cdots &I_\nu(\xi_n^*) \\ 
\vdots              &      &  \vdots \\   
\xi_1^{*\,n-m-1}I_\nu^{(n-m-1)}(\xi_1^*) 
&\cdots& 
\xi_n^{*\,n-m-1}I_\nu^{(n-m-1)}(\xi_n^*)\\ 
\end{array} 
\!\!\!\right) 
\nn\\ 
&& 
\label{Z1qqweak} 
\eea 
suppressing the symmetry factor $m!$ as well as taking the transpose of the 
matrix. At the last step the $m$ integrations over the $\la_k$ have been taken 
inside the rows 
of the determinant. Eq. (\ref{Z1qqweak}) generalizes the 
results obtained in \cite{SV03} 
for an equal number $m=n$ of quarks and conjugate  
anti-quarks 
of degenerate, complex conjugate mass $\hat{m}=z\one_n$ and  
$\hat{n^*}=z^*\one_n$ each.  
In \cite{SV03} the partition function is given as well for $n\geq m$, for  
real degenerate masses $\hat{m}=x\one_m$ and $\hat{n}^*=y\one_n$.  
 
\indent 
 
We can now compare to the second model by taking the weak nonhermiticity limit 
of eq. (\ref{Z2qqN}). To achieve this we first multiply all powers $m_f^\nu$  
into the first $m$ rows and all powers $n_g^{*\,\nu}$ into all $n$ columns  
of the determinant ${\cal B}(m_i,n^*_j)$.  
The large-$N$ limit of the norms $h_k$ eq. (\ref{norm}) and the bare kernel  
eq. (\ref{kernel}) have already been taken in \cite{A02}. 
Ignoring all factors of $N$ the product of norms will lead to  
\be 
\lim_{N\to\infty\,\tau\to1} \prod_{k=N}^{N+m-1}h_k\ \sim \ 
\al^m\ . 
\label{limnorm} 
\ee 
The weak nonhermiticity limit of the bare kernel is given by  
\cite{A02} 
\bea 
&&\lim_{N\to\infty\,\tau\to1} 
m_f^\nu n^{*\,\nu}_g \kappa_{N+m}(-m_f^2, -n^{*\,2}_g)  
\ \sim \ \frac{1}{\al} \int_0^1d\la\la \mbox{e}^{-\al^2\la^2}I_\nu(\la\zeta_f) 
I_\nu(\la\xi_g^*) \ , 
\label{limkernel} 
\eea 
where we have continued to negative arguments of the kernel. We have used 
again the value $\qq=1/\sqrt{2}$ in eq. (\ref{microweak}) corresponding to our 
model.  
Since all matrix elements of kernels have different arguments there is no 
degeneracy in this part of the determinant.  
Multiplying with eq.  
(\ref{limnorm}) cancels all inverse powers of $\al$ from the kernel inside the 
determinant. 
The part of the determinant containing only polynomials 
$\tilde{P}_k^{(\nu)}(-n^{*\,2})$ can be dealt with as previously, using 
eqs. (\ref{weaktau}) and (\ref{Laguerreasymp}). First we take out all powers 
of $\tau$, leading to  
\be 
\lim_{N\to\infty\,\tau\to1} \tau^{(n-m)(2N+n+m-1)\frac12} \ \sim \ 
\exp\left[ -\frac12 (n-m)\al^2\right]\ . 
\ee 
The rows of polynomials become degenerate after taking the limit eq.  
(\ref{Laguerreasymp}), leading again to differentiations as in 
eq. (\ref{Z2weak}).  
Performing the manipulations, we arrive at the following result 
\bea 
&&{\cal Z}^{(N_f=m+n,\nu)}_{II}(\al;\{\zeta_f\}_m,\{\xi_g^*\}_n) 
\arrowvert_{\mbox{weak}} \ \sim 
\frac{\exp\left[-\frac12(n-m)\al^2\right]}{ 
\Delta_m(\zeta^2)\Delta_n(\xi^{*\,2})} \nn\\ 
&& 
\times\det 
\left( 
\begin{array}{ccc} 
\int_0^1d\la\la 
\,\mbox{e}^{-\al^2\la^2}I_\nu(\la\zeta_1) I_\nu(\la\xi_1^*) 
&\cdots& 
\int_0^1d\la\la 
\,\mbox{e}^{-\al^2\la^2}I_\nu(\la\zeta_1)I_\nu(\la\xi_n^*)\\  
\vdots              &      &  \vdots \\   
\int_0^1d\la\la 
\,\mbox{e}^{-\al^2\la^2}I_\nu(\la\zeta_m) I_\nu(\la\xi_1^*) 
& \cdots&  
\int_0^1d\la\la 
\,\mbox{e}^{-\al^2\la^2}I_\nu(\la\zeta_m)I_\nu(\la\xi_n^*)\\  
I_\nu(\xi_1^*) &\cdots &I_\nu(\xi_n^*) \\ 
\vdots              &      &  \vdots \\   
\xi_1^{*\,n-m-1}I_\nu^{(n-m-1)}(\xi_1^*) 
&\cdots& 
\xi_n^{*\,n-m-1}I_\nu^{(n-m-1)}(\xi_n^*)\\ 
\end{array} 
\right)\!. 
\nn\\ 
&& 
\label{Z2qqweak} 
\eea 
The equivalence of the two partition functions can be established as follows.  
If we multiply eq. (\ref{Z2qqweak}) by the overall constant  
$\exp[+\frac14 N_f\al^2]$, with $N_f=n+m$, and then identify  
\be 
(1-\tau^2)\ =\ 4\qq^2\mu^2 
\label{mutau2} 
\ee 
or equivalently $\al=2\tilde{\al}$ the two partition functions 
(\ref{Z1qqweak}) and (\ref{Z2qqweak}) agree. We conclude that  
both models can be mapped onto each other as a function of rescaled masses  
and rescaled nonhermiticity parameter, for  
an arbitrary number of quarks and conjugate anti-quarks of different,  
nondegenerate masses. In the present case the dependence on $\al$ is much 
less trivial  
as it enters all the integrals inside the determinant. 
We are not able to provide a simple explanation 
for the fact that the two identifications  
eqs. (\ref{mutau1}) and (\ref{mutau2}) differ only by a factor of two. 
However, our main point here is that there exists a mapping of parameters
making the two partition functions the same. We will find yet another mapping 
of parameters in the limit of strong nonhermiticity below.

\subsection{The strong nonhermiticity limit}\label{strong} 
 
The limit of strong nonhermiticity is defined by keeping $\mu^2$ or 
$\tau\in[0,1]$ fixed, independent of $N$ when performing the large-N limit.
Furthermore, the quark masses (and the complex Dirac eigenvalues) have to be 
rescaled with a different power in $N$ \cite{A02,A03}\footnote{For the 
  constant proportionality factor we have kept $\qq$ as in the weak limit, 
  although the macroscopic spectral density will in general no longer be 
  constant but rather depend on $\mu$. For a $\mu-$dependent rescaling 
  we refer to the discussion after 
  eq. (\ref{Z2strongmu}) and after eq. (\ref{Z2qqstrong}).} 
\bea 
\zeta_f &\equiv& \sqrt{2N}\qq m_f \ , f=1,\ldots,m \nn\\ 
\xi_g^* &\equiv& \sqrt{2N}\qq n_g^*\ , g=1,\ldots,n\ . 
\label{microstrong} 
\eea 
In this limit complex eigenvalues of the Dirac operator fill in a truly two
dimensional domain in the complex plane and the corresponding
spectral density is nonvanishing there.  
With these definitions at hand we can again make a saddle point approximation 
for the partition functions in question.

We start with the first model eq. (\ref{Z1ev}) which contains only quarks.  
Due to the rescaling of the masses the arguments of the 
Bessel functions now become large. But even when replacing them with their 
asymptotic value, $I_\nu(x)\sim \exp[x]/\sqrt{x}$, they will not 
contribute to the saddle point, as the argument is of order 
${\cal O}(\sqrt{N})$ and is small compared to terms ${\cal O}(N)$ 
in the exponent for
radial variables. The saddle point is thus given by conditions
\be 
\frac{2r_k}{r_k^2-\mu^2}\ -\ 2r_k\qq^2 \ =\ 0,\ \  k=1,\ldots,N_f\ . 
\label{strongSPeq} 
\ee 
and we choose the positive value  
\be 
\hat{r}\arrowvert_{\mbox{sp}}\ =\ \one_{N_f} \sqrt{\qq^{-2}+\mu^2}\ . 
\label{strongSP} 
\ee 
as the relevant solution.
For a more detailed analysis, in particular, on the connection of the 
saddle point solution at zero to the phase transition  
we refer to section \ref{phase} below. 
Taking the usual limit of a degenerate matrix at  
$\hat{r}\arrowvert_{\mbox{sp}}\sim\one_{N_f}$ 
we arrive at 
\bea 
{\cal Z}^{(N_f,\nu)}_{I}(\mu;\{\zeta_f\})\arrowvert_{\mbox{strong}}&\sim&  
\exp[-NN_f(1+\qq^2\mu^2)] \  
\exp\left[-\frac12\sum_{k=1}^{N_f}\zeta_k^2\right]\nn\\ 
&\times&\frac{\det_{k,l=1,\ldots,N_f} 
\left[\zeta^{k-1}_l I_{\nu}^{(k-1)}\left(\sqrt{2N}\zeta_{l} 
\sqrt{1+\qq^2\mu^2}\right) 
\right]}{\Delta_{N_f}(\zeta^2)}\ . 
\label{Z1strong} 
\eea 
 
\indent 
 
Now we treat the second model by looking at eq. (\ref{Z2N}) at finite-$N$.  
The strong nonhermiticity limit is most easily taken here by using an 
integral representation of the Laguerre polynomials, following \cite{A03} for 
$N_f=1$, 
\bea 
x^\nu L_N^{(\nu)}(x^2) &\sim& \mbox{e}^{x^2}\int_0^\infty  
ds\ \mbox{e}^{-Ns} s^{N+\frac{\nu}{2}} J_\nu(2\sqrt{sN}x) \nn\\ 
&\sim& \mbox{e}^{x^2}\mbox{e}^{-N} J_\nu(2\sqrt{N}x) \ . 
\label{OPintrep} 
\eea 
At the second step we have made a saddle point approximation, 
taking into account that 
$x^2=-Nm^2/(2\tau)$ is fixed and finite. Using the rescaling of the masses 
eq. (\ref{microstrong}) with the chosen value $\qq=1/\sqrt{2}$,  
corresponding to $\sqrt{N}m=\zeta$, and
taking care of the degeneracy of the determinant we obtain 
\bea 
{\cal Z}^{(N_f,\nu)}_{II}(\tau;\{\zeta_f\})\arrowvert_{\mbox{strong}}&\sim&  
\exp[-NN_f]\ \tau^{N_f(N+\frac14 (N_f-1)+\frac12\nu)}\  
\exp\left[-\frac{1}{2\tau}\sum_{k=1}^{N_f}\zeta_k^2\right] 
\nn\\ 
&\times&\frac{\det_{k,l=1,\ldots,N_f} 
\left[\zeta^{k-1}_l I_{\nu}^{(k-1)}\left(\sqrt{2N}\zeta_{l}\tau^{-\frac12} 
\right) 
\right]}{\Delta_{N_f}(\zeta^2)}\ . 
\label{Z2strongtau} 
\eea 
Using now the relation eq. (\ref{mutau1}), 
\be 
\tau\ =\ \sqrt{1-2\qq^2\mu^2} 
\ee  
we see that the $\mu-$dependence does not match the previous
case in general, unless we expand 
for small $\mu^2$. To the leading order we obtain  
\bea 
{\cal Z}^{(N_f,\nu)}_{II}(\mu;\{\zeta_f\})\arrowvert_{\mbox{strong}}&\sim& \! 
\exp[-NN_f(1+\qq^2\mu^2)+{\cal O}(\mu^4)] 
\exp\!\left[-\frac{1}{2}(1+\qq^2\mu^2+{\cal O}(\mu^4)) 
\sum_{k=1}^{N_f}\zeta_k^2\right]\nn\\ 
&\times&\!\frac{\det_{k,l=1,\ldots,N_f} 
\left[\zeta^{k-1}_l I_{\nu}^{(k-1)}\left(\sqrt{2N}\zeta_{l} 
(1+\frac12\qq^2\mu^2+{\cal O}(\mu^4)) 
\right) 
\right]}{\Delta_{N_f}(\zeta^2)}\ . 
\label{Z2strongmu} 
\eea 
The two expressions eqs. (\ref{Z1strong}) and (\ref{Z2strongmu}) 
match to the leading order terms in the determinant and in the first factor.  
However, the coupling between masses $\zeta_k$ 
and $\mu$ introduced in the second model  
in the second exponential prefactor does not have an analogue in the first 
model. Here the matching only holds to terms of zeroth order. 
 
It is worth mentioning that even apart from the $\mu-$dependent exponential  
suppression factor $\exp[-NN_f\qq^2\mu^2]$ a proper large-$N$ limit of the 
partition functions in terms of rescaled masses $\zeta_k$ does not exist.  
The Bessel functions still depend on $\sqrt{N}$ in the argument. In the 
asymptotic limit however, the exponential suppression wins.  
Our microscopic rescaling of the masses eq. (\ref{microstrong}) cannot be 
modified to achieve an $N$-independent result. 
Such a rescaling is, in fact, dictated by corresponding
rescaling of the complex eigenvalues \cite{A02} necessary to obtain a smooth 
limiting eigenvalue  
correlation functions. 
This type of rescaling in $N$ at the regime of strong nonhermiticity was 
also found earlier in models without chiral symmetry \cite{A01}.  
 
One also may wonder if it is possible to match the two partition functions 
eqs. (\ref{Z1strong}) and (\ref{Z2strongtau})  beyond an 
expansion in $\mu$, by introducing a different, $\mu$- and $\tau$-dependent 
rescaling of the masses in eq.  (\ref{microstrong}) respectively.  
However, a little thought shows that due to essentially
different $\tau$- and $\mu$-dependence of the masses
of the two models such a procedure is impossible.  
 
\indent

In order to take the large-$N$ limit at strong nonhermiticity of the 
first model with quarks and conjugate anti-quarks we go back to the 
expressions for finite-$N$ of the partition function, 
eqs. (\ref{Z1qqQm}) and (\ref{Z1qqUR}).  At the regime of
weak nonhermiticity it was crucial to expand the determinant 
$\det[\cdots]^N$ to be able to perform
exactly the unitary group integrals arising from the parameterization of 
$Q$.  However, at the regime of strong nonhermiticity such an expansion
is no longer  possible with respect to the parameter $\mu$ 
since the latter no longer scales with $N$.  For 
the sake of simplicity we will restrict ourselves to real quark and 
conjugate anti-quarks masses, $M=M^\dagger$, eq. (\ref{Z1qqUR}). In general 
the radial and unitary degrees of freedom no longer decouple. 
However, if we assume that the permutation symmetry makes all
radial coordinates $r_k$ to take the same saddle point value as it was the 
case previously, see eq. (\ref{weakSP}), that means
\be 
\hat{r}\arrowvert_{\mbox{sp}}\ =\ r_{\mbox{sp}}\one_{N_f}\ , 
\label{SPansatz} 
\ee 
with the value $r_{\mbox{sp}}$ to be determined, the integral  
eq. (\ref{Z1qqUR}) simplifies considerably. 
The unitary matrix $V$ drops out  from the integrand and we obtain 
\bea 
{\cal Z}^{(N_f,\nu)}_I(\mu;M)&\sim& \mbox{e}^{-N\qq^2\Tr\, M^2}  
\mbox{e}^{-N\qq^2N_f r_{\mbox{sp}}^2} \  
\int dU \det[U^\dagger]^\nu 
\nn\\ 
&\times&  
\exp\left[ 
{N\qq^2r_{\mbox{sp}}\Tr M (U+U^\dagger)} 
\right] 
\det\left[r_{\mbox{sp}}^2\one_{N_f} 
-\mu^2 U^\dagger\Sigma_3 U\Sigma_3  
\right]^N . 
\label{Z1qqURSP} 
\eea 
This expression can be regarded as the effective partition function at 
the regime of strong 
nonhermiticity, and serves as generalization of 
the effective Lagrangian from chiral  
perturbation theory \cite{TV}, eq. (\ref{Z1qqUVSP}) which was valid 
at weak nonhermiticity.  
The two expressions only agree after
expanding to the first order in $\mu^2$ (for real masses). 
Note that while in the chiral Lagrangian picture higher powers terms  
$(\mu^2U^\dagger\Sigma_3 U\Sigma_3)^k$  
can be excluded from  power counting in $U$, they 
cannot be excluded in the matrix model. We will see that in order to describe
a phase transition to the symmetric phase it will be important to keep  
all powers in $\mu$, see section \ref{phase} below, also cf.\cite{HJV}.

In the following we also restrict ourselves to an equal number of quarks and 
their conjugate partners, $n=m$. The reason is that in the general  
case $n>m$ one naturally expects 
a result to be of the mixed form (compare  eq. (\ref{Z2qqN}) or  
eq. (\ref{Z1qqweak}), containing both limiting "kernels"
and "polynomials"\footnote{For the interpretation of the partition function  
in terms of kernels and polynomials we refer to section 
  \ref{univ}.}. However, we have just seen that the quantities corresponding 
to the polynomials do not have a proper large-$N$ limit. Since we are looking 
for partition functions that do possess  
such a limit at strong nonhermiticity (as 
we will find for the second model below)  
we choose $n=m$ to ensure that the result contains only kernel terms. 
 
In eq. (\ref{Z1qqURSP}) we can again employ the parameterization 
eqs. (\ref{Upar}) and (\ref{Lapar}), where it is instructive to first look at  
the determinant alone.  
We obtain 
\bea 
\det\left[r_{\mbox{sp}}^2\one_{N_f}  
-\mu^2 U^\dagger\Sigma_3 U\Sigma_3 \right]^N &=& 
\det\left( 
\begin{array}{cc} 
r_{\mbox{sp}}^2\one_n - \mu^2v_1(2\hat{\lambda}^2-\one_n)v_1^\dagger 
& 2 \mu^2v_1\hat{\lambda}\sqrt{\one_n-\hat{\lambda}^2})v_2^\dagger\\ 
-2 \mu^2v_2\hat{\lambda}\sqrt{\one_n-\hat{\lambda}^2})v_1^\dagger 
&r_{\mbox{sp}}^2\one_n-\mu^2v_2(2\hat{\lambda}^2-\one_n)v_2^\dagger\\ 
\end{array} 
\right)^N\nn\\ 
&=& \det\left[r_{\mbox{sp}}^4\one_{n}  
-2r_{\mbox{sp}}^2\mu^2(2\hat{\lambda}^2-\one_n)+\mu^4\one_n\right]^N 
\nn\\ 
&=&(r_{\mbox{sp}}^2+\mu^2)^{2nN} 
\prod_{i=1}^n\left(1-\frac{4r_{\mbox{sp}}^2\mu^2}{(r_{\mbox{sp}}^2+\mu^2)^2}
\lambda_i^2 \right)^N , 
\label{rmudet} 
\eea 
where all angular-variable dependence has dropped out.  
The integrals over the unitary subgroups $u_{1,2}$ and $v_{1,2}$  
of the mass dependent exponential  
$\exp[N\qq^2r_{\mbox{sp}}\Tr M (U+U^\dagger)]$ can be performed as in the 
weak nonhermiticity limit eq. (\ref{Z1qqint}), using the integral 
eq. (\ref{BTint}). In fact we can literally repeat all the following 
simplifying steps 
there after eq. (\ref{Z1qqdet^2}), leading to  
\bea 
&&{\cal Z}^{(N_f,\nu)}_I(\mu;\{\zeta_f\}_n,\{\xi_g\}_n) \ \sim\  
\frac{1}{\Delta_n(\zeta^2)\Delta_n(\xi^2)}\  
\mbox{e}^{-\frac12\sum_{k=1}^n(\zeta_k^2+\xi_k^2)}   
\ r_{\mbox{sp}}^{N_f(\nu+1)}  
\mbox{e}^{-N\qq^2N_f r_{\mbox{sp}}^2} \ (r_{\mbox{sp}}^2+\mu^2)^{2nN} 
\nn\\ 
&&\times 
\det_{i,j=1,\ldots,n}\left[ 
\int_0^1 d\la \la 
\left(1-\frac{4r_{\mbox{sp}}^2\mu^2}{(r_{\mbox{sp}}^2+\mu^2)^2}\lambda^2  
\right)^N  
I_\nu\left(\sqrt{2N}\qq r_{\mbox{sp}}\la \zeta_i\right) 
I_\nu\left(\sqrt{2N}\qq r_{\mbox{sp}}\la \xi_j\right) 
\right]. 
\label{Z1qqLa} 
\eea 
Here we have also inserted the microscopic scaling of the masses 
eq. (\ref{microstrong}). We still have to determine the saddle point value 
$r_{\mbox{sp}}$ and take the large-$N$ limit. In the latter the  
integral can be computed by making the change of variables  
$s= 4r_{\mbox{sp}}^2\mu^2\,N\la^2/(r_{\mbox{sp}}^2+\mu^2)^2$ and using that  
$\lim_{N\to\infty}(1-s/N)^N=\exp[-s]$. We also employ the following integral, 
\be 
\int_0^\infty ds \mbox{e}^{-s}  
J_\nu\left(\sqrt{2s}\zeta\right)J_\nu\left(\sqrt{2s}\xi\right) 
\ =\ \exp\left[ -\frac12(\zeta^2+\xi^2) \right] I_\nu(\zeta\xi) \ , 
\label{Iint} 
\ee 
after analytically continuing in the masses\footnote{The integral in 
  eq. (\ref{Z1qqLa}) is also convergent after change of variables and taking 
  the large-$N$ limit as it stands, without continuing to imaginary masses.}.  
Taking out common factors of the determinant we arrive at  
\bea 
&&{\cal Z}^{(N_f,\nu)}_I(\mu;\{\zeta_f\}_n,\{\xi_g\}_n) \ \sim\  
\frac{1}{\Delta_n(\zeta^2)\Delta_n(\xi^2)}\  
\mbox{e}^{-\frac12\sum_{k=1}^n(\zeta_k^2+\xi_k^2)}   
\mbox{e}^{-N\qq^2N_f r_{\mbox{sp}}^2} \ (r_{\mbox{sp}}^2+\mu^2)^{2nN} 
\nn\\ 
&&\times 
\left(\frac{(r_{\mbox{sp}}^2+\mu^2)^2}{4Nr_{\mbox{sp}}^2\mu^2}\right)^n 
\mbox{e}^{\frac12\qq^2(r_{\mbox{sp}}^2+\mu^2)^2 
\sum_{k=1}^n\frac{(\zeta_k^2+\xi_k^2)}{4\mu^2}}  
\det_{i,j=1,\ldots,n}\left[ 
I_\nu\left(\frac{\qq^2 (r_{\mbox{sp}}^2+\mu^2)^2\zeta_i\xi_j}{4\mu^2} 
\right) 
\right]. 
\label{Z1qqrsp} 
\eea 
The value for $r_{\mbox{sp}}$ can finally be read off as
\be 
(r_{\mbox{sp}}^2+\mu^2)\ =\ \qq^{-2} \ , 
\ee 
which allows to simplify the final result down to  
\bea 
{\cal Z}^{(N_f,\nu)}_I(\mu;\{\zeta_f\}_n,\{\xi_g\}_n) &\sim&  
\frac{1}{\Delta_n(\zeta^2)\Delta_n(\xi^2)} \frac{1}{(4\qq^2\mu^2)^n}\  
\mbox{e}^{-2nN(\qq^2\mu^2-1)} \label{Z1qqfinalstr}\\ 
&&\times 
\exp\left[ 
\frac12(1-4\qq^2\mu^2) 
\sum_{k=1}^n\frac{(\zeta_k^2+\xi_k^2)}{4\qq^2\mu^2}\right]   
\det_{i,j=1,\ldots,n}\left[ 
I_\nu\left(\frac{\zeta_i\xi_j}{4\qq^2\mu^2} 
\right) 
\right]. 
\nn 
\eea 
Our conclusion is therefore that the large-$N$ limit of the 
partition function (as a functions of the masses)
with an equal number of quarks and conjugate anti-quarks is well-defined,  
in contrast  
to that for quarks alone, eq. (\ref{Z1strong}). 
The $\mu$-dependence can be  
almost entirely absorbed by redefining the rescaling of the masses 
eq. (\ref{microstrong}) to $\zeta_f\to\zeta_f/(2\qq\mu)$ and 
$\xi_g\to\xi_g/(2\qq\mu)$, giving: 
\bea 
{\cal Z}^{(N_f,\nu)}_I(\mu;\{\zeta_f\}_n,\{\xi_g\}_n) &\sim&  
\frac{\mbox{e}^{-2nN(\qq^2\mu^2-1)} }{\Delta_n(\zeta^2)\Delta_n(\xi^2)}  
\frac{1}{(4\qq^2\mu^2)^n}\  
\mbox{e}^{\frac12(1-4\qq^2\mu^2) 
\sum_{k=1}^n(\zeta_k^2+\xi_k^2)} 
\!\det_{i,j=1,\ldots,n}\left[ 
I_\nu\left(\zeta_i\xi_j\right) 
\right]. 
\nn\label{Z1qqstr2n} 
\eea

\indent 
 
The strong nonhermiticity limit of the second model with quarks and conjugate 
quarks is again performed easily, without being restricted to real masses.  
Looking at the expression eq. (\ref{Z2qqN}) we 
need the limiting normalization factors, kernel and polynomials. The latter 
have been analyzed by us already, and we start with the normalization factors. 
Omitting all constant factors  
we obtain  
\be 
\prod_{k=N}^{N+m-1} h_k \ \sim\ (f^{(\nu)}(\tau))^m\ , 
\label{normfactor} 
\ee 
which contains all $\tau-$dependence. 
Before taking the strong nonhermiticity limit for the kernel (and polynomials) 
we multiply  
all the prefactors $m_f^\nu$ and $n_g^{*\,\nu}$  
into the determinant $\det[{\cal B}(m_i,n^*_j)]$ in eq. (\ref{Z2qqN}),  
as in the weak limit before. 
The asymptotic kernel has been already evaluated in \cite{A02}, reading  
\bea 
m^\nu n^{*\,\nu} \kappa_N(-m^2,-n^{*\,2}) &\sim&  
\frac{1}{(1-\tau^2)} \frac{1}{f^{(\nu)}(\tau)}\  
\exp\left[\frac{\tau}{2(1-\tau^2)}(\zeta^2+\xi^{*\,2})\right] 
I_\nu\left(\frac{\zeta \xi^{*}}{1-\tau^2}\right) . 
\label{strongkernel} 
\eea 
Multiplying all factors of $f^{(\nu)}(\tau)$ from eq. (\ref{normfactor}) into 
the first $m$ rows of the determinant cancels the $f^{(\nu)}(\tau)$-dependence 
of the kernel. Using eq. (\ref{OPintrep}) for the asymptotic polynomials   
results in the full expression for the partition function reading 
\bea 
&&{\cal Z}^{(N_f=m+n,\nu)}_{II}(\tau;\{\zeta_f\}_m,\{\xi_g^*\}_n) 
\arrowvert_{\mbox{strong}} \ \sim 
\frac{\mbox{e}^{-N(n-m)}}{ 
\Delta_m(\zeta^2)\Delta_n(\xi^{*\,2})} \ \frac{1}{(1-\tau^2)^m}\  
\tau^{(n-m)(N+\frac14(n+3m-1+2\nu))}\nn\\ 
&& 
\times\det 
\left( 
\begin{array}{ccc} 
\mbox{e}^{\frac{\tau}{2(1-\tau^2)}(\xi_1^{*\,2}+\zeta_1^2)} 
I_\nu(\frac{\xi_1^*\zeta_1}{1-\tau^2})  
&\cdots& 
\mbox{e}^{\frac{\tau}{2(1-\tau^2)}(\xi_n^{*\,2}+\zeta_1^2)} 
I_\nu(\frac{\xi_n^*\zeta_1}{1-\tau^2})\\  
\vdots              &      &  \vdots \\   
\mbox{e}^{\frac{\tau}{2(1-\tau^2)}(\xi_1^{*\,2}+\zeta_m^2)} 
I_\nu(\frac{\xi_1^*\zeta_m}{1-\tau^2})  
& \cdots&  
\mbox{e}^{\frac{\tau}{2(1-\tau^2)}(\xi_n^{*\,2}+\zeta_m^2)} 
I_\nu(\frac{\xi_n^*\zeta_m}{1-\tau^2})\\  
\mbox{e}^{\frac{-\xi_1^{*\,2}}{2\tau}}I_\nu(\sqrt{2N}\xi_1^*\tau^{-\frac12})  
&\cdots & 
\mbox{e}^{\frac{-\xi_n^{*\,2}}{2\tau}}I_\nu(\sqrt{2N}\xi_n^*\tau^{-\frac12})\\ 
\vdots              &      &  \vdots \\   
\mbox{e}^{\frac{-\xi_1^{*\,2}}{2\tau}} 
\xi_1^{*\,n-m-1}I_\nu^{(n-m-1)}(\sqrt{2N}\xi_1^*\tau^{-\frac12}) 
&\cdots& 
\mbox{e}^{\frac{-\xi_n^{*\,2}}{2\tau}} 
\xi_n^{*\,n-m-1}I_\nu^{(n-m-1)}(\sqrt{2N}\xi_n^*\tau^{-\frac12})\\ 
\end{array} 
\right)\!.  
\nn\\ 
&& 
\label{Z2qqstrong} 
\eea 
In order to compare with eq. (\ref{Z1qqstr2n}) we look at  
the particular case of equal number of quarks and conjugate anti-quarks, 
$n=m$,   
\bea 
{\cal Z}^{(N_f=2n,\nu)}_{II}(\tau;\{\zeta_f\}_n,\{\xi_g^*\}_n) 
\arrowvert_{\mbox{strong}} &\sim& 
\frac{1}{ 
\Delta_n(\zeta^2)\Delta_n(\xi^{*\,2})} \ \frac{1}{(1-\tau^2)^n}\  
\exp\left[{\frac{\tau}{2} 
\sum_{k=1}^n\frac{\xi_k^{*\,2}+\zeta_k^2}{1-\tau^2}}\right] 
\nn\\ 
&& 
\times\det_{i,j=1,\ldots,n} 
\left[I_\nu\left(\frac{\xi_i^*\zeta_j}{1-\tau^2}\right)\right] . 
\label{Z2qqstrong2n} 
\eea 
This expression has a finite limit at large $N$ 
as a function of $\tau$ and the masses.  
We note that here the masses always appear in the same $\tau-$dependent 
combination $1/\sqrt{1-\tau^2}$. If we define a $\tau-$dependent microscopic 
rescaling $m\sqrt{2N/(1-\tau^2)}=\zeta$ instead of eq. (\ref{microstrong}) 
that square-root factor can be absorbed:  
\bea 
{\cal Z}^{(N_f=2n,\nu)}_{II}(\tau;\{\zeta_f\}_n,\{\xi_g^*\}_n) 
\arrowvert_{\mbox{strong}} &\sim& 
\frac{1}{ 
\Delta_n(\zeta^2)\Delta_n(\xi^{*\,2})} \ \frac{1}{(1-\tau^2)^n}\  
\mbox{e}^{\frac{\tau}{2} 
\sum_{k=1}^n(\xi_k^{*\,2}+\zeta_k^2)} 
\det_{i,j=1,\ldots,n} 
\left[I_\nu\left(\xi_i^*\zeta_j\right)\right] .\nn\\ 
&& 
\label{Z2qqstr2n} 
\eea 
We note that   
$\rho(0)=\frac{1}{2\pi(1-\tau^2)}$ is the constant macroscopic 
density corresponding to uniform filling of an ellipse. 
The scaling defined above is very reminiscent to the 
scaling on the real line used to get read of the mean level spacing.   
The two partition functions eq.  (\ref{Z1qqstr2n}) and (\ref{Z2qqstr2n})  
can now be identified upon the mapping  
\be 
\tau=1-4\qq^2\mu^2\ . 
\label{mutau3} 
\ee 
It is yet different from the two previous mappings, the main point of our 
analysis being the very existence of 
a correspondence between the two functions in terms of their arguments.

\subsection{Universality}\label{univ} 
 
After having demonstrated several cases of
agreement between the two partition
function with a given flavors content  
let us use those results for discussing
the issue of universality of the random matrix theory in the complex plane.  
For matrix models with eigenvalues on the real line universality is  
a well established principle, both on the base of heuristic, physical arguments
\cite{ADMN,DN,AF} as well on firm mathematical grounds (see \cite{KV,SF2} and 
references therein;  we mainly discuss 
works relevant for QCD applications).  
On the other hand, for complex eigenvalues the subject has so far 
attracted only little attention, apart from the discussion in \cite{A02II}. 
As we already have mentioned several times, it is important 
to distinguish between 
the limits of weak and strong nonhermiticity \cite{FKS}. 
As is well known, the eigenvalue correlations in the weak nonhermiticity
limit interpolate between those typical for real
eigenvalues  and those known for complex eigenvalues.  
More precisely, sending the rescaled nonhermiticity parameter $\al$ 
from eq. (\ref{weaklim}) to $0$ or $\infty$
allows to recover correlation functions typical for eigenvalues of 
Hermitian or, respectively, complex
strongly nonhermitian (Ginibre-like) matrices.  
This was explicitly checked in several cases, as for example in  
\cite{FKS} for the unitary (UE) and in  
\cite{A02} for the chiral unitary ensemble (chUE).  
 
In that sense the weak nonhermiticity limit is closely
related to real, universal 
correlations, and it is natural to expect it should maintain 
at least some universality features. 
In \cite{A02II} a certain class of deformations of 
measures of the complex UE were studied with the conclusion, that the 
asymptotic polynomials, kernels and correlation functions remain universal.  
 
It therefore came as a surprise that for the complex extension of the chUE 
two slightly different results  were obtained 
for the microscopic spectral density. 
Indeed, the second model eq. (\ref{Z2}) with complex eigenvalues 
was solved in the paper \cite{A02} by analytical
continuation of the Laguerre polynomials, and a comparison 
to quenched QCD lattice data with chemical potential 
confirmed the obtained results 
\cite{AWII}.  
On the other hand, the original model eq. (\ref{Z1}) based on a 
random matrix representation as a starting point
was solved recently for the microscopic spectral density in the 
weak limit, using an exact replica approach, see \cite{SV03}. 
The obtained density, although 
having a very similar structure, was slightly different,
and the difference was confirmed by simulating numerically the 
underlying random matrix 
model, eq. (\ref{Z1}). 
We remark however, that for the range of parameters $\mu$ used in the  
analysis of the lattice data  
\cite{AWII} the difference was too small to be detected.   
In view of the {\it agreement} between partition functions of the two 
models found in the preceding sections and in \cite{A03} 
we will try to shed some more 
light on the origin of the discrepancy,
and thus on the issue of universality.

In the random matrix theory several different objects 
can be tested for the property of universality.
Let us start with recalling some basic facts for models that 
can be solved with the technique of orthogonal polynomials, such as 
eq. (\ref{Z2}) or the chUE. The solution for the eigenvalue correlation 
functions can be written as  
\bea 
\rho_k(z_1,\ldots,z_k) &=& \det_{i,j=1,\ldots,k}[K_N(z_i,z_j^*)]\ =\  
\prod_{l=1}^k w(z_l,z_l^*) \ \det_{i,j=1,\ldots,k}[\kappa_N(z_i,z_j^*)] 
\nn\\ 
&=& \prod_{l=1}^k w(z_l,z_l^*) \  
\det_{i,j=1,\ldots,k}\left[\sum_{k=0}^{N-1}h_k^{-1} \tilde{P}_k^{(\nu)}(z_i) 
\tilde{P}_k^{(\nu)}(z_j^*)\right]\ , 
\label{rhos} 
\eea 
where we use the notation introduced in eqs. (\ref{kernel}) and 
(\ref{OP}).  

It is evident, that three different objects can 
be analyzed from the point of view of asymptotic universality in the 
large-$N$ limit: 
the polynomials $\tilde{P}_N^{(\nu)}(z)$, the bare kernel 
$\kappa_N(z_i,z_j^*)$ 
and the weight function $w(z,z^*)$ itself. All these objects have a 
direct relation to partition functions as we will see below.  
For the universality of the spectral correlations eq. (\ref{rhos}) we need 
that both the weight and the bare kernel are universal, as the full kernel is 
given by  
$K_N(z_i,z_j^*)= [w(z_i,z_i^*)w(z_j,z_j^*)]^{\frac12}\kappa_N(z_i,z_j^*)$.  
We also note that for real eigenvalues the universality of the bare kernel 
follows directly from that of the polynomials\footnote{We note 
  however, that the universality of the norms $h_k$ is a separate issue, see 
  e.g. in \cite{ADMN}.} 
due to the Christoffel-Darboux formula  
\be 
\kappa_N(x,y)\ =\ h_{N-1}^{-1}\  
\frac{\tilde{P}_N(x)\tilde{P}_{N-1}(y)-\tilde{P}_N(y)\tilde{P}_{N-1}(x)}{x-y} 
\ . 
\label{CD} 
\ee 
For orthogonal polynomials in the complex plane this relation does not hold 
and thus the universality of polynomials does not necessarily imply the 
universality of the bare kernel, and vise versa.  
 
There exists an alternative method for 
computing spectral correlation functions from
working with characteristic polynomials. An advantage of the method 
is that it retains its validity also 
in a general case when orthogonal polynomials are unavailable, 
for example,  when the partition 
function cannot be represented in terms of eigenvalues. In the simplest case 
the resolvent $G(z)$ can be generated from differentiating 
a single ratio of characteristic 
polynomials, 
\be 
G(z)\ \equiv\ \left\langle \Tr \frac{1}{z-H}\right\rangle_N\ =\  
\left.{\partial_z} \left\langle \frac{\det(z-H)}{\det(u-H)} 
\right\rangle_N\right|_{u=z} \ . 
\label{resolvent} 
\ee 
Here $H=H^\dagger$ denotes a hermitian random matrix of size $N\times N$ 
averaged over a Gaussian 
or more general weight function. Knowing the resolvent, the spectral density 
follows by taking the discontinuity along the support or the 
antiholomorphic derivative, for real or complex eigenvalues, respectively.  
For getting access to 
higher order correlation functions more ratios of characteristic 
polynomials can be used as source terms.  
 
Apart from being a generator for spectral correlation functions 
characteristic polynomials can be regarded as interesting 
objects in their own right, and it is therefore natural to
ask about their universality.  
In fact, it is possible to 
relate characteristic polynomials to orthogonal
polynomials in such models where the latter are known. 
 First of all, the expectation value
of a single characteristic polynomial 
directly gives the orthogonal polynomial for the corresponding model
in monic normalization: 
\be 
\left\langle \det(z-H) \right\rangle_N \ =\ \tilde{P}_N(z)\ . 
\label{ch-OP} 
\ee 
Second, the expectation value 
of the product of two characteristic polynomials yields 
directly the bare kernel 
(see e.g. in \cite{PZJ}) 
\be 
 \left\langle \det(z-H) \det(u-H) \right\rangle_N \ =\  
h_N\ \kappa_{N+1}(z,u) \ . 
\label{ch-ker} 
\ee 
Third, for real eigenvalues the expressions involving
the inverse of a characteristic polynomial are given in terms of the  
Cauchy transform of the orthogonal polynomials, as was recently discovered
 in \cite{FS3}, and further developed in \cite{SF2,BDS,FA,AF}. For example:
\be 
 \left\langle \frac{1}{\det(z-H)}\right\rangle_N \ =\  
\frac{1}{2\pi i} \int dx \frac{w(x)}{z-x}\tilde{P}_N(x)\ \equiv\  
\vartheta(z)\ ,
\label{ch-intOP} 
\ee 
where to ensure that the object is well-defined, 
the poles have to be suitably avoided by giving an imaginary part to  
$z$, $\im(z)\neq 0$. Various formulas expressing arbitrary products 
\cite{BH,FW,MN,A01,AVIII} and arbitrary ratios \cite{FS3,SF2,BDS} of 
characteristic 
polynomials in terms of orthogonal polynomials, 
their Cauchy transforms and bare kernels  
containing one or both of these have been proved recently.  
 
A natural question which immediately arises is whether 
the universality also holds for arbitrary ratios of  
characteristic polynomials, especially in view of 
their relation to orthogonal 
polynomials and bare kernels just mentioned. 
We would like to point out that because of the presence of the Cauchy 
transform the universality of such objects is in general 
not a simple consequence  
of the known universality of the kernels and polynomials.  
For the unitary ensembles with real eigenvalues this question has been 
completely answered. For the standard UE the universality was rigorously 
proved in 
\cite{SF2}.  
Arbitrary ratios of characteristic polynomials for the chUE 
have been computed 
in \cite{FS1,FA,SV02} and proved to be universal in various regimes
in \cite{AF,V}. 
Let us mention that all what we have said immediately applies to the  
matrix model partition functions of QCD, as the insertion of massive flavors 
is nothing else than the insertion of $N_f$ characteristic polynomials.  
In fact, from the point of view of massive partition functions 
the universality
of arbitrary products has been proved previously in \cite{DN} 
for the UE and 
chUE.  
 
For characteristic polynomials of matrices with complex eigenvalues much less 
is known. First of all we have to distinguish between the characteristic  
polynomial and its complex conjugate, just as we have distinguished between 
quarks and conjugate anti-quarks in our previous considerations.  
For arbitrary products of characteristic polynomials and a different number of 
complex conjugate characteristic polynomials an expression in terms of 
polynomials and bare kernels has been derived in \cite{AVIII}  
(see eq. (\ref{Z2qqN}) which we have already used).  
From what has been said in this section  
it is clear, that using products alone we will  
not be able to deduce the universality complex eigenvalue correlations.  
 
We can thus conclude the following. We have studied two different models, 
eqs. (\ref{Z1}) and (\ref{Z2}), which are apriori not the same, in particular 
as the former does not admit a simple eigenvalue representation.  
We have found that in the large-$N$ limit at weak nonhermiticity  
the two partition functions agree for an arbitrary and different number of 
quarks and conjugate anti-quarks, showing that they both belong to the same 
universality class. This gives a strong argument in favor of   
universality
for arbitrary products of characteristic polynomials
in the regime of weak
nonhermiticity  . For complex 
models without chiral symmetry this follows from \cite{A02II} by proving 
universality of the kernel and polynomials there.  
 
At strong nonhermiticity we found that the partition functions 
of the two models can be mapped onto each other, 
provided we consider an equal number of quarks and conjugate anti-quarks. 
This correspondence indicates possible universality of some strongly 
nonhermitian partition functions. 
Another argument could be that for the second model  
its weakly nonhermitian (universal)  
partition function can be matched to the strongly  
nonhermitian one by taking the limit $\al\to\infty$. 
However, we found that two partition functions  
with a general flavor content,  for example containing  
only quarks, disagree in general. 
Despite the fact that they have a very similar structure,
only the leading order term in expansion with respect to $\mu$
can be put in correspondence.  
This indicated that universality at 
strong nonhermiticity regime may be more subtle.  
When discussing such a disagreement 
some caution has to be added. As we have already remarked,  
a large-$N$ limit of the partition functions with general flavor content 
as functions of the masses does not exist, properly speaking.   

Let us come back to the question of universality of correlation functions at 
weak nonhermiticity. In \cite{SV03} the corresponding microscopic spectral 
density was computed for the first model eq. (\ref{Z1}) at $\nu=0$.  
There, an exact 
replica method was used, that  expresses the density in terms of a product  
and a ratio of characteristic polynomials, 
\bea 
\rho_{I}\arrowvert_{\mbox{weak}}(\zeta) &=& \frac12 |\zeta|^2 
{\cal Z}^{(N_f=-1-1,0)}_{I}(\mu;\zeta,\zeta^*) \  
{\cal Z}^{(N_f=1+1,0)}_{I}(\mu;\zeta,\zeta^*)\ , 
\label{weakrhoII} 
\eea 
written in terms of the rescaled variable $\zeta=2N\qq z$. 
Here by a negative number $N_f=-2$ of flavors we indicate the corresponding  
number of inverse powers of determinants $m=n=1$ in eq. (\ref{Z1qq}). 
The latter objects in the present context 
are also frequently called the bosonic partition functions.  
Let us compare this expression to the result of \cite{A02} 
\bea 
\rho_{II}\arrowvert_{\mbox{weak}}(\zeta) &=& \lim_{N\to\infty\tau\to 1} 
w(z,z^*)\ \kappa_N(z,z^*) \nn\\ 
&=& \lim_{N\to\infty\,\tau\to 1} 
w(z,z^*)\ h_{N-1}^{-1} {\cal Z}^{(N_f=1+1,\nu)}_{II}(\tau;z,z^*) \ , 
\label{weakrhoI} 
\eea 
where in the second step we have used eq.  (\ref{Z2qqN}) for $n=m=1$,  
corresponding to eq. (\ref{ch-ker}) in the complex plane. 
As we know, the two fermionic partition functions, that is corresponding to
positive value $N_f=2$, ${\cal Z}^{(N_f,\nu)}_{I,II}$ 
agree for any value of $\nu$. We therefore can compare 
the weight function of the 
second model in the weak 
nonhermiticity limit with 
$\frac12|\zeta|^2{\cal Z}^{(N_f=-1-1,0)}_{II}$ that plays 
the same role.  
While for the weight factor we have  
\bea 
\lim_{N\to\infty\,\tau\to1} w(z,z^*)\ h_N^{-1} &=& \frac{1}{\al} 
|\zeta|^{2\nu+1}\exp\left[-\frac{1}{\al^2}(\im \zeta)^2 
\right] 
\label{weakweight} 
\eea 
in terms of the rescaled variable, for the bosonic partition functions   
taken from \cite{SV03} it follows\footnote{Note that in \cite{SV03} the 
matrices are  chosen to be antihermitian.}  
\bea 
|\zeta|^2{\cal Z}^{(N_f=-1-1,0)}_{I}(\al;\zeta,\zeta^*)  
&\sim&  
\frac{|\zeta|^2}{\al^2}\exp\left[\frac{1}{4\al^2}(\zeta^2+\zeta^{*\,2})\right] 
K_0\left(\frac{|\zeta|^2}{2\al^2}\right)\ . 
\label{Z1bos} 
\eea 
In the limit of small $\al$ or large $|\zeta|^2$ we can expand the $K$-Bessel 
function $K_0(x)\sim\mbox{e}^{-x}/\sqrt{x}$ and both expressions  
eqs. (\ref{weakweight}) and (\ref{Z1bos})  will
coincide for $\nu=0$. However in general they disagree, leaving different 
possibilities for the universality of spectral correlation functions at weak 
nonhermiticity. We are left with two possibilities, one being 
that  the models are in different universality classes 
for the spectral correlations, despite agreeing for all correlations of 
products of  
characteristic polynomials. Or, alternatively, 
there is no spectral universality at weak 
nonhermiticity at all, in the sense that such a universality 
only holds for characteristic polynomials and not 
for the weight function itself.  
 
To further illustrate these possibilities  
let us point out a major difference from the Hermitian large-$N$ limit. 
At zero chemical potential the weight function corresponding  
$\exp[-NV(x)]$, with $V(x)$ an even polynomial, is reduced to a trivial unity 
factor in  
the microscopic scaling limit $Nx=const$ . For that reason the distinction 
between bare and full kernel becomes immaterial in that limit. Therefore the 
agreement between the two models for the product of two characteristic  
polynomials 
leading to the kernel, eq. (\ref{ch-ker}) implies the same
agreement for the 
spectral correlations.   
However, at weak nonhermiticity this is not the case any longer. 
As we see in 
eq. (\ref{weakweight}) the weight function remains different from unity in 
that limit. In fact it has to be a function of the imaginary part 
$\im(\zeta)$  and the rescaled chemical potential $\al$, such that it 
reduces to a 
$\delta-$function in $\im(\zeta)$ in  
the limit $\al\to 0$\footnote{We would like to mention that in the analysis 
  \cite{A02II} only such deformations of the weight functions were studied 
  which keep a Gaussian representation of the $\delta-$function. Such 
  deformations still do enter in the macroscopic density $\rho(0)$.}.  
Otherwise the model would not 
reduce to the chUE as a model of QCD \cite{SV93}. 
 
Finally  
we would like to mention that there exists a matrix model \cite{James}  
different from the 
two models studies in this paper. It is given in terms of two 
independent complex matrices and can be diagonalized to give a complex 
eigenvalue representation. The weight function of the eigenvalues of  
this model is precisely given by eq. (\ref{Z1bos}).  
The solution of the model \cite{James} by the method of orthogonal 
polynomials provides an very interesting third alternative model for QCD  
with chemical potential, having simultaneously a matrix and complex eigenvalue 
representation. While the microscopic density at weak nonhermiticity  
agrees with that found in  \cite{SV03} the strong limit has not been  
studied so far.  
 
Another approach to the open problem of universality would be to  
compute the bosonic partition function in the  
second model ${\cal Z}^{(N_f=-2,0)}_{II}$. Thus it remains to be seen if the  
universality found for products breaks  
down for inverse powers of characteristic polynomials.

 
\sect{Phase transitions}\label{phase} 
 
The aim of this section is to study the chiral  
phase transition at a critical value $\mu_c$.  
Since the second model eq. (\ref{Z2}) is always in the broken phase 
\cite{A02}, with a constant macroscopic density on an ellipse for all allowed 
values of $\tau\in[0,1]$, we will only study the first model eq. (\ref{Z1}).  
Furthermore, we will restrict ourselves to the presence of quarks alone. This 
is mainly because only in this case a very  
compact expression eq. (\ref{Z1ev}) is 
available (compared to eqs. (\ref{Z1qqQm}) or (\ref{Z1qqUR})). 
 
The same model has already been studied for one flavor $N_f=1$ in great 
detail \cite{HJV}. The virtue of the expression  eq. (\ref{Z1ev}) is that most 
of this analysis carries over to several flavors.  
An extended version of the model eq. (\ref{Z1}) including temperature has been 
studied to predict the phase diagram of QCD for two light flavors 
\cite{HJSSV}.   
However, in order to be able to solve 
the model rather strong assumption had to be made on the nature of the 
saddle point.  
One of the main goals of our present analysis is to see 
if such a consideration for zero temperature can 
be checked and further extended, with an additional number 
of not necessarily 
degenerate flavors included into the model.  
More recently the model with temperature and chemical potential  
\cite{HJSSV}  
has been extended to study the different effect of baryon and isospin  
chemical potential for two flavors \cite{KTV}.  
The latter corresponds to having a pair of a quark 
and its conjugate, and the authors find a doubling of the 
critical line as compared with \cite{HJSSV}.   
 
Here there is an important difference 
from the previous section in the large-$N$ 
limit. We are interested in finding discontinuities 
of the partition function  
eq. (\ref{Z1ev}) as a function of masses and chemical potential $\mu$ 
when $N$ becomes large. In 
contrast to the weak and strong nonhermiticity limit considered before, 
we therefore will 
not assume any scaling of the masses $m_f$ with $N$. This will modify  
the respective saddle point eqs. (\ref{weakSP}) and (\ref{strongSP}) 
as the Bessel function will now make a nontrivial contribution to it. 
It introduces a mass 
dependence into the saddle point that makes an analysis of the effect of  
light versus heavy flavors possible. 
 
Before taking the saddle point limit we further compactify the expression 
eq. (\ref{Z1ev}), by multiplying the Vandermonde determinant of the radial 
coordinates, $\Delta_{N_f}(r^2)=\det_{i,j}[r_i^{2(j-1)}]$, with the determinant
of Bessel functions. The resulting matrix elements $(i,j)$ of the single 
determinant read $\sum_{k=1}^{N_f}r_k^{2(j-1)}I_\nu(m_i r_k)$. Due to the 
symmetry of the partition function under permutations of the $r_k$ 
we can follow the 
same steps as described after eq. (\ref{Z1qqdet^2}). This reduces the 
determinant to $N_f!$ times a simpler determinant with  
elements $r_j^{2(j-1)}I_\nu(m_i r_j)$. We can now take the integration over 
each $dr_j$ into the $j-$th column, and obtain up to the symmetry factor  
$ N_f!$ 
\bea 
{\cal Z}^{(N_f,\nu)}_I(\mu;\{m_f\}) &=& 
\mbox{e}^{-N\qq^2\sum_{f=1}^{N_f}m_f^2}\ \frac{1}{\Delta_{N_f}(m^2)} \ 
\label{Z1evcompact}\\ 
&&\times\det_{i,j=1,\ldots,N_f} \left[ 
\int_0^\infty  
dr\, r^{\nu+1+2(j-1)} (r^2-\mu^2)^N \mbox{e}^{-N\qq^2 r^2}  
I_\nu(2N\qq^2m_i\, r)  
\right]\ .  
\nn 
\eea 
In this form valid at finite-$N$ the partition function looks exactly  
as a determinant over finite-$N$ partition functions of a {\it single} flavor, 
apart for the difference between the index of the Bessel function $\nu$ 
and the different power in $r$ to $(\nu+1 +2(j-1))$.  
At the saddle point this difference 
will be of course subleading and allows for the analysis of discontinuities of 
the $N_f$ flavors in terms of a one-flavor partition function elaborated in 
\cite{HJV}.  We can now evaluate the saddle points of the integrals 
individually. Taking the asymptotic limit for the Bessel function  
$I_\nu(x)\sim$ e$^x/\sqrt{x}$ we obtain for each row 
\be 
\frac{r}{r^2-\mu^2}\ -\ r\qq^2 \ +\ \qq^2 m_i\ =\ 0\ \ ,\  
i=1,\ldots,N_f\ . 
\label{phaseSP} 
\ee 
In order to lift the resulting degeneracy of the determinant we have to 
differentiate the Bessel functions as usual. Furthermore, we note that since 
it is only their asymptotic exponential behavior that enters the saddle point 
equation we interchange the differentiation and the saddle point procedures.  
The result is  
\bea 
{\cal Z}^{(N_f,\nu)}_I(\mu;\{m_f\}) &=& 
\frac{\mbox{e}^{-N\qq^2\sum_{f=1}^{N_f}m_f^2}}{\Delta_{N_f}(m^2)}  
\det_{i,j=1,\ldots,N_f} \left[ 
m_i^{j-1} \partial_{m_i}^{j-1} z_I^{(N_f=1,\nu)}(\mu;m_i)\arrowvert_{\mbox{sp}}
\right],  
\label{Z1phase} 
\eea 
where we have defined the partition function with its trivial exponential  
mass dependence removed, 
\be 
 z_I(\mu;m)\ \equiv\ \mbox{e}^{+N\qq^2m^2} 
\ {\cal Z}^{(N_f=1,\nu)}_I(\mu;m)\ . 
\label{reducedZ} 
\ee 
The large-$N$ partition function is thus given by a determinant of 
(differentiated) single flavor partition functions at their saddle point 
value. This enables us to draw some general conclusions.

A first order phase transition occurs if the first logarithmic 
derivative of the partition function is discontinuous at some 
value. We look for a discontinuity with respect to $\mu$ so we define 
the quark number density 
\be 
n_q\ \equiv\ \frac{1}{N_f} \partial_\mu \ln 
{\cal Z}^{(N_f,\nu)}_I(\mu;\{m_f\}) \ . 
\label{nqdef} 
\ee 
Applying this to our result eq. (\ref{Z1phase}),  
\bea 
n_q &=&  
\Tr\ln\left[\frac{\partial_\mu A}{A}\right]  
\ ,\ \  
(A)_{ij}\equiv m_i^{j-1} \partial_{m_i}^{j-1}  
z_I(\mu;m_i)\arrowvert_{\mbox{sp}}\ , 
\label{nqN_f} 
\eea 
we observe that a discontinuity occurs whenever an individual differentiated 
matrix element becomes singular. Thus, the phase transition arises when  
the one-flavor partition function ${\cal Z}_I^{(1,\nu)}(\mu;m_i)$ 
with the smallest value $\mu_c$ becomes discontinuous. Since $\mu_c$ is a 
functions of the mass $m_i$, we have to compare the $m_i$-dependence of the 
saddle point equation (\ref{phaseSP}) with its corresponding  
value $\mu_c(m_i)$. We will find that it 
is always the smallest mass (which may 
be zero) that has the smallest value of $\mu_c$ and thus drives the 
transition. 
 
The analysis of the saddle point solution and the corresponding partition 
function for $N_f=1$ has been made already in great detail in \cite{HJV}, and 
we follow it closely. 
The saddle point equation (\ref{phaseSP}) is of third order and thus may have 
up to three real solution. We begin with the simplest, massless case: 
\be 
\frac{r}{r^2-\mu^2}\ -\ r\qq^2 \ =\ 0\ \ \Rightarrow\ \  
r\arrowvert_{\mbox{sp}}\ =\ \left\{ 
\begin{array}{lll} 
0                      &=r_r(m=0)& \ \mbox{restored}\\ 
+\sqrt{\qq^{-2}+\mu^2} &=r_b(m=0)& \ \mbox{broken}\\ 
-\sqrt{\qq^{-2}+\mu^2} & & \ \notin\ [0,\infty)\\ 
\end{array}\right.\ . 
\label{m=0SP} 
\ee 
It is easy to see that the solution with $r\arrowvert_{\mbox{sp}}>0$ belongs  
to the broken phase $(b)$  
with an exponentially suppressed partition functions,  
while the solution with $r\arrowvert_{\mbox{sp}}=0$ corresponds to the 
restored phase (r) \cite{HJV}. 
The negative solution is rejected being outside the 
integration domain.  
If we switch on a mass $m$ the signature of the saddle point will remain 
the same\footnote{This can be seen from the discriminant  
for all real values of the masses.},  
with solutions being $0< r_r(m)<r_b(m)$ as we will see below.  
 
The equation that determines the critical value $\mu_c$ is given by the 
requirement of partition functions at two competing 
saddle points being equal, 
\be 
(r_b^2-\mu^2)\mbox{e}^{\qq^2(2m r_b- r_b^2)} \  
=\ (\mu^2-r_r^2)\mbox{e}^{\qq^2(2m r_r- r_r^2)}\ . 
\label{muc} 
\ee 
In the massless case $m=0$ this leads, after inserting the solutions 
(\ref{m=0SP}),  to the following equation for the 
critical line:
\be 
1+\qq^2\mu_c^2 + \ln[\qq^2\mu_c^2]\ =\ 0\ ,\ \ \Rightarrow\  
\qq\mu_c \approx 0,527\ldots\  
\ \ , 
\label{mucmin} 
\ee 
where we have given the approximate numerical value for its real solution  
(see also fig \ref{muplot} below). 
 
Let us now determine how this value shifts if we include 
a small mass. We first present a perturbative analysis to the leading order in 
the mass\footnote{A similar analysis was made in \cite{KTV} for $N_f=2$ with a 
  different chemical potential for each flavor.} and then come back to  
the full solution below.  
First we determine the shift of the saddle point solution due to the  
mas term, 
\bea 
 r_r(m) -  r_r(0) &\equiv& \delta_r(m)\ =\  
\frac{m\qq^2\mu^2}{1+\qq^2\mu^2} 
\nn\\ 
 r_b(m) -  r_b(0)  &\equiv& \delta_b(m)\ =\  
\frac{m}{2(1+\qq^2\mu^2)}\ , 
\label{shift} 
\eea 
where we retained only the leading order linear behavior  
in $m$.

\begin{figure}[-h] 
\centerline{ 
\epsfig{figure=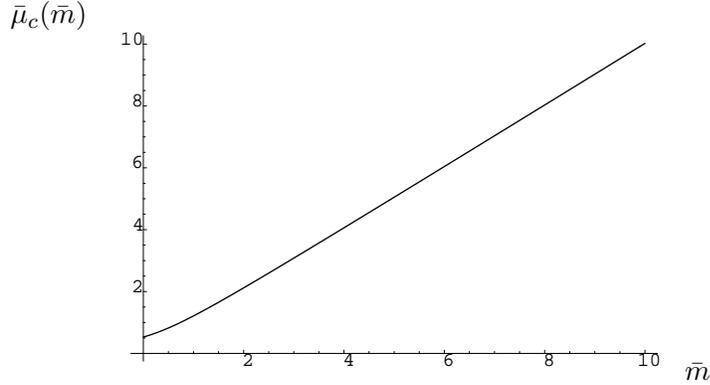,width=18pc} 
\put(5,0){$\bar{m}$} 
\put(-250,135){$\bar{\mu}_c(\bar{m})$} 
} 
\caption{ 
The critical potential as a function of mass, both given in units of $\qq$.  
The critical value at $\bar{m}=0$ is $\bar{\mu}_c(0)=0.527\ldots$ . 
}\label{muplot} 
\end{figure}

Next we compute the 
resulting shift in the critical line, determining $\mu_c(m)$: 
\be 
1+\qq^2\mu_c^2 + \ln[\qq^2\mu_c^2]\ -\ 2m\qq^2\sqrt{\qq^{-2}+\mu_c^2}  
+{\cal O}(m^2)\ =\ 0\ . 
\label{mucshift} 
\ee 

To the leading linear order in $m$ both $\delta_{r,b}$ dropped out.  
We thus obtain for the shift in $\mu_c^2$ defined as 
\be 
\mu_c^2(m) \ -\ \mu_c^2(0)ß \equiv\ \gamma(m) \ , 
\label{gammadef} 
\ee 
the positive quantity 
\be 
\gamma(m)\ =\ \frac{2\qq m \qq^2\mu_c(0)}{\sqrt{1+\qq^2\mu_c(0)}}\ +\  
{\cal  O}(m^2) \ , 
\label{gamma} 
\ee 
after inserting it  
into eq. (\ref{mucshift}). It is again given to leading order 
only. From this result we can deduce the following. If we have massless and 
massive flavors present in the partition function eq. (\ref{Z1phase})  
the critical value for the massless single flavor partition functions will be 
reached before that of the massive flavors, as $\mu_c(0)<\mu_c(m\neq0)$.  
Second, if only massive flavors are present it is the critical $\mu_c(m)$ of 
the lightest flavor which is reached first, as for $m_1<m_2$ holds  
$\mu_c^2(m_1)<\mu_c^2(m_2)$.  
In both cases it is the lightest or zero mass that 
triggers the phase transition. Since it is of first order for a single flavor 
\cite{HJV} it is thus first order for any combination of massive and massless 
flavors, as follows from eq. (\ref{nqN_f}). 
Moreover, for each different mass there is separate transition, leading in 
principle to a sequence of transitions\footnote{We wish to emphasize that in 
  general the flavor dependence of matrix models is too weak. In QCD the 
  sign of the $\beta-$function changes for sufficiently many flavors, and the 
  finite temperature transition changes from second to first order when 
  changing from 2 to 3 flavors, which is not reproduced from the matrix model 
  \cite{HJSSV}. For that reason our conclusion should be taken seriously 
  for small $N_f$ only.}.  
A similar 
feature of two distinct first order lines was found in \cite{KTV} for two 
flavors.  
 
We have also convinced ourselves that the monotonic behavior of the critical 
chemical potential $\mu_c(m)$ persists beyond the linear 
approximation, eqs. (\ref{gamma}) and  (\ref{gammadef}).  
This can be done by inserting the two positive  
real solutions for the third order saddle point equation into the condition 
eq. (\ref{muc}) and solving it numerically. If we measure all quantities in 
units of $\qq$, $\bar{r}=\qq r$,  $\bar{\mu}=\qq \mu$ and $\bar{m}=\qq m$ 
(or set $\qq=1$) we have  
\bea 
\bar{r}_r &=& \frac23 \sqrt{3(1+\bar{\mu}^2)+ \bar{m}^2}\cos[\varphi/3+4\pi/3] 
+\frac{\bar{m}}{3} \nn\\ 
\bar{r}_b &=& \frac23 \sqrt{3(1+\bar{\mu}^2)+ \bar{m}^2}\cos[\varphi/3] 
+\frac{\bar{m}}{3} \nn\\ 
&&\cos[\varphi]\ =\  
\frac{\bar{m}(\bar{m}^2-9\bar{\mu}^2+\frac92)}{[3(1+\bar{\mu}^2)+ 
\bar{m}^2]^{\frac32}} \ , 
\label{rsolution} 
\eea 
as the two positive solutions of the saddle point equation. 
At $m=0$ ($\varphi=\pi/2$) the first solution $\bar{r}_r$ vanishes as  
it should. 
We have plotted above the numerical solution of eq. (\ref{muc}),  
after inserting into it the full solution eq. (\ref{rsolution}). 
It can be seen that $\bar{\mu}_c(\bar{m})$ is a monotonous function  
beyond the regime of small masses {$\bar{m}$.

In order to check our findings we have examined numerically the partition  
function for two flavors $N_f=2$ with two nondegenerate masses, in  
the representation of eq. (\ref{Z1evcompact}). We have chosen the parameters 
$m_1=1$ and $m_2=2$ in units of $\qq$ and $\nu=0$.  
The resulting logarithm  
of the partition function is plotted above for three  
different values of $N$  as a function of $\mu$ (figure \ref{Zplot}). 
\begin{figure}[-t] 
\centerline{ 
\epsfig{figure=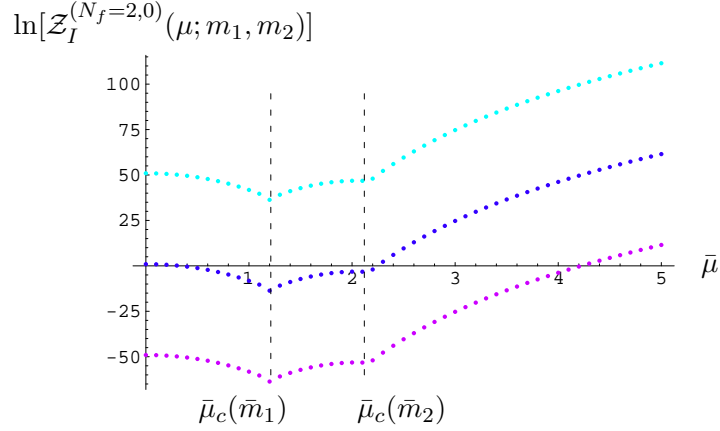,width=18pc} 
\put(10,45){$\bar{\mu}$} 
\put(-250,135){$\ln[{\cal Z}_{I}^{(N_f=2,0)}(\mu;m_1,m_2)]$} 
\put(-180,-10){$\bar{\mu}_c(\bar{m}_1)$} 
\put(-120,-10){$\bar{\mu}_c(\bar{m}_2)$} 
} 
\caption{ 
The logarithm of the partition function for N=10 (top), N=20 (middle), 
and N=30 (lower line). The vertical lines indicate the positions  
of the of the corresponding critical values $\mu_c(m_{1,2})$. 
}\label{Zplot} 
\end{figure} 
Even for the smallest value of $N$ the points where the derivative of the  
partition function and thus the free energy becomes  
discontinuous are very well  
visible. The positions coincide with the critical values of $\mu$ for each  
mass scale, $\bar{\mu}_c(\bar{m}_1)=1.2119\ldots$ 
and $\bar{\mu}_c(\bar{m}_2)=2.119\ldots$. (from fig. \ref{muplot}).

After having made a general statement we give an explicit example for the  
partition function eq. (\ref{Z1phase})  
below and above the transition.  
Here we can again make use of analysis 
\cite{HJV} where 
we briefly repeat some of the results.  
The single flavor partition function (modulo its mass prefactor 
eq. (\ref{reducedZ})) can be conveniently written as
\bea 
z_I(\mu;m) &=& 
\int_0^{\mu^2} dr\, r(r^2-\mu^2)^N \mbox{e}^{-N r^2} I_0(2Nm\, r)  
\ +\  
\int_{\mu^2}^\infty  
dr\, r(r^2-\mu^2)^N \mbox{e}^{-Nr^2} I_0(2Nm\, r) \nn\\ 
&\equiv& z_r(\mu;m) \ +\  z_b(\mu;m) \ . 
\label{zrbdef} 
\eea 
We used \cite{HJV} that the restored symmetry solution satisfies 
$r_r\in[0,\mu^2]$ 
and the broken symmetry solution is $r_b\in[\mu^2,\infty)$ (see also 
eq. (\ref{rsolution})). Here and in the following we set $\qq=1$ and 
$\nu=0$ as it is done in \cite{HJV}. The two different solutions can be written
as a double sum at finite-$N$, 
\bea 
z_r(\mu;m) &=& \mu^{2(N+1)}\mbox{e}^{-n\mu^2}  
\sum_{l=0}^\infty\sum_{s=0}^{\infty}\frac{(N+s)!}{l!s!(N+l+s+1)!} 
(Nm^2)^l(N\mu^2)^{l+s}\ , 
\nn\\ 
z_b(\mu;m)&=&\frac{1}{N^{N+1}} \mbox{e}^{-n\mu^2}  
\sum_{l=0}^\infty\sum_{s=0}^{l}\frac{(N+s)!}{l!s!(l-s)!}(Nm^2)^l(N\mu^2)^{l-s} 
\label{zrbN}\ . 
\eea 
These formulas easily generate the massless partition function  
and its derivatives, $z_{r,b}^{(l)}(\mu;m=0)$, as we need them in 
eq. (\ref{Z1phase}). We note that only even derivatives contribute and  
give the examples necessary for $N_f=2$ massless flavors. 
They read in the restored phase  
\bea 
z_r(\mu;m=0) &\sim& \mu^{2(N+1)} \nn\\ 
z_r''(\mu;m)\arrowvert_{m=0} &\sim& \mu^{2(N+1)}2N\mu^2\ , 
\eea 
and in the broken phase 
\bea 
z_b(\mu;m=0) &\sim& \mbox{e}^{-N\mu^2} \nn\\ 
z_b''(\mu;m)\arrowvert_{m=0}  &\sim& \mbox{e}^{-N\mu^2}  
2N^2(1+\mu^2+1/N)\ . 
\eea 
Taking the limit of two degenerate, massless flavors in eq. (\ref{Z1phase}) 
we obtain for large $N$  
\bea 
{\cal Z}^{(N_f=2,\nu=0)}_I(\mu;m=0) &=& 
\det\left( 
\begin{array}{cc} 
z_I(\mu;0) & 0\\ 
0                  & z_I''(\mu;0)\\ 
\end{array} 
\right)\sim\left\{ 
\begin{array}{cc} 
\mu^{4N}                   & \mbox{restored}\\ 
\mbox{e}^{-2N\mu^2}(1+\mu^2)& \mbox{broken}\\ 
\end{array} 
\right. 
\label{Z1phaseN_f2}\ .  
\eea 
This compares to the single flavor result \cite{HJV} 
\bea 
{\cal Z}^{(N_f=1,\nu=0)}_I(\mu;m=0)  
\ \sim\ \left\{ 
\begin{array}{cc} 
\mu^{2N}                   & \mbox{restored}\\ 
\mbox{e}^{-N\mu^2}& \mbox{broken}\\ 
\end{array} 
\right. 
\label{Z1phaseN_f1}\ .  
\eea 
We see that the behavior is very similar, with the quark number density in 
the two flavor case being twice as high. 
 
As we have mentioned already there is more than one transition in the 
presence of several mass scales. In particular this opens the possibility to 
have for example two transitions for two massless flavors  
and one massive flavor, 
as it is often used in a simple model for QCD.  
For $\mu_c(0)<\mu<\mu_c(m)$, the massless building blocks 
are already in the restored phase while the massive 
ones are still in the broken phase. This is an intermediate regime before 
full symmetry restoration.  
It would be very interesting to use the exact solution of our model as a  
further testing ground for lattice algorithms as for example in \cite{AANV}. 
For the discussion of phenomenological consequences 
we refer to \cite{KTV} where a similar phenomenon is observed.

 
\sect{Summary}\label{disc} 
 
We have computed and compared  
two different matrix model partition functions for QCD with chemical potential.
While the first model has only a matrix representation and no eigenvalue
representation the situation is the reverse for the second model.  
We give very compact, new  expressions for finite-$N$  
for the first model in terms of integrals over the radial coordinates of the  
underlying sigma-model-like representation. They hold for an arbitrary number  
of quarks or both quarks and conjugate anti-quarks.  
By taking the  
large-$N$ limit these expressions are then 
compared to the corresponding results from the second model 
expressed in terms of  
orthogonal polynomials in the complex plane and their kernel.  
 
In the limit of weak nonhermiticity we find a complete agreement, 
with a different identification of the nonhermiticity parameters  
$\mu$ and $\tau$ and mass parameters
in the cases of only quarks present, or both quarks and conjugate  
anti-quarks.  
  
The matching between the two models
in the limit of strong nonhermiticity holds  
only for an equal number of quarks and conjugate  
anti-quarks. For such a matching the masses have  
to be rescaled with respect to the $\mu$- or $\tau$-dependent macroscopic  
density proportional to the level spacing. For a generic number of  
quarks and conjugate anti-quarks, including the case of only quarks, 
a mapping  
can be achieved when perturbatively expanding in small chemical  
potential $\mu$, despite the very similar structure of the two partition  
functions. However, in this case a proper large-$N$ limit of the two partition 
functions does not exist.

From these finding we concluded that at weak nonhermiticity  
and also in a special case at strong nonhermiticity the  
expressions for the fermionic partition functions are universal. 
The issue of universality of eigenvalue correlation functions is left open, 
as it involves also the computation of bosonic partition functions.  
 
In the last part we have  
investigated the phase structure of the first matrix model, 
which is known to possess a first order phase transition for a single  
or several degenerate flavours (the second model is always in the symmetry
broken  phase by construction).  
We have analyzed the model for an arbitrary  
number of nondegenerate quark flavours, exploiting the compact expressions  
derived for finite-$N$.  
In the large-$N$ limit they give the multi-flavour partition function  
as a determinant of single flavor partition functions of different  
masses, and their derivatives.  
Consequently for each nondegenerate flavor 
a separate first order  
transition occurs, where the critical $\mu$ increases monotonically in mass. 
We have checked our findings numerically for  
two flavors with nonvanishing, nondegenerate masses. 
 
It would be very interesting to compute also the corresponding bosonic  
partition functions and compare them, in particular in the light  
of their relation to complex eigenvalue correlations and the related question  
of universality.

\indent 
 
\noindent 
\underline{Acknowledgments}: 
We wish to thank  
B. Schlittgen, K. Splittorff, and D. Toublan  
for valuable discussions.  
We especially thank J. Osborn for sharing with us results prior to 
publication and J. Verbaarschot for helpful comments on the manuscript.  
We acknowledge a kind hospitality extended to us at the ECT*, Trento 
where this project 
was initiated.  
This work was supported by  
a Heisenberg fellowship of the Deutsche Forschungsgemeinschaft (GA),  
by a Brunel University Vice Chancellor Grant (YVF) and by  
the European network on ``Discrete Random 
Geometries'' HPRN-CT-1999-00161 EUROGRID (GV).

\begin{appendix} 
 
\sect{The sigma-model representation of the partition function  
}\label{sigmarep} 
 
Here we derive the sigma-model representation for both the partition 
functions eqs. (\ref{Z2}) and (\ref{Z1}). The latter can be obtained by 
omitting all conjugate flavors. 
To this aim we first unify notation, by writing eq. (\ref{Z1qq}) as 
\be 
{\cal Z}^{(N_f,\nu)}_I(\mu;\{\om_f\}) 
= (-1)^{n(2N+\nu)} 
\int d\Phi d\Phi^\dagger  
\prod^{N_f}_{f=1}  
\det\left( 
\begin{array}{c} 
\om_f\one_N                       \ \ \ \     i\Phi + \mu_f\tilde{\one}_N\\ 
\\ 
i\Phi^\dagger + \mu_f\tilde{\one}_{N}^\dagger  \ \ \ \    \om_f\one_{N+\nu}\\ 
\end{array} 
\right)  
\exp[-N\qq^2\Tr\Phi\Phi^\dagger], 
\label{Z1qqA} 
\ee 
denoting the complex masses with 
\be 
\om_f\ \equiv\ \left\{ 
\begin{array}{ll} 
+m_f & f=1,\ldots,m\\ 
-n_f^* & f=m+1,\ldots,N_f=m+n\\ 
\end{array} 
\right. \ , 
\label{om} 
\ee 
as well as choosing signs for the chemical potentials  
$\mu_f=+\mu$ for $f=1,\ldots,m$, and $\mu_f=-\mu$ for $f=m+1,\ldots,N_f=m+n$.  
After introducing two sets of complex Grassmann vectors $\chi_A$ and $\chi_B$ 
of size $N$ and $N+\nu$ respectively, we can express the determinant 
in eq. (\ref{Z1}) as  
\bea 
{\cal Z}^{(N_f,\nu)}_I(\mu;\{\om_f\})&\sim& \int d^2\chi_A d^2\chi_B\  
\mbox{e}^{ 
-\sum_{f=1}^{N_f} \left(\om_f (\chi_{A\,f}^\dagger \chi_{A\,f} 
+\chi_{B\,f}^\dagger \chi_{B\,f}) 
\ +\ \mu_f(\chi_{A\,f}^\dagger\tilde{\one}_N  
\chi_{B\,f} +\chi_{B\,f}^\dagger\tilde{\one}_N^\dagger \chi_{A\,f}) 
\right) 
} 
\nn\\ 
&&\times\left\langle  
\mbox{e}^{-i\sum_{f=1}^{N_f} (\chi_{A\,f}^\dagger \Phi\chi_{B\,f} 
+\chi_{B\,f}^\dagger \Phi^\dagger \chi_{A\,f})} 
\right\rangle_\Phi . 
\label{step1} 
\eea 
We define the expectation value over the ensemble
amounting to the integration over matrices $\Phi$ as  
\bea 
\!\!\!\!\left\langle  
\mbox{e}^{-i\sum_{f=1}^{N_f} (\chi_{A\,f}^\dagger \Phi\chi_{B\,f} 
+\chi_{B\,f}^\dagger \Phi^\dagger \chi_{A\,f})} 
\right\rangle_\Phi &\equiv& 
\int d\Phi d\Phi^\dagger 
\mbox{e}^{-i\sum_{f=1}^{N_f} (\chi_{A\,f}^\dagger \Phi\chi_{B\,f} 
+\chi_{B\,f}^\dagger \Phi^\dagger \chi_{A\,f})} 
\ \mbox{e}^{-N\qq^2\Tr\Phi^\dagger\Phi} \!\! . 
\label{expPhi} 
\eea 
It can be further rewritten as  
\bea 
\left\langle  
\mbox{e}^{-i\sum_{f=1}^{N_f} (\chi_{A\,f}^\dagger \Phi\chi_{B\,f} 
+\chi_{B\,f}^\dagger \Phi^\dagger \chi_{A\,f})} 
\right\rangle_\Phi &=&  
\left\langle  
\mbox{e}^{+i\,\Tr( \Phi\sum_{f=1}^{N_f} \chi_{B\,f}\otimes\chi_{A\,f}^\dagger 
\ +\  \Phi^\dagger\sum_{f=1}^{N_f}\chi_{A\,f}\otimes \chi_{B\,f}^\dagger)} 
\right\rangle_\Phi 
\nn\\ 
&=& \exp\left[-\frac{1}{N\qq^2}\sum_{f,g=1}^{N_f}\Tr 
(\chi_{A\,f}\otimes\chi_{B\,f}^\dagger) 
(\chi_{B\,g}\otimes \chi_{A\,g}^\dagger)\right] 
\nn\\ 
&=&\exp\left[-\frac{1}{N\qq^2}\sum_{f,g=1}^{N_f} 
(\chi_{A\,g}^\dagger\chi_{A\,f})(\chi_{B\,f}^\dagger\chi_{B\,g})\right] 
\nn\\ 
&=& \int dQ dQ^\dagger \mbox{e}^{\,\sum_{f,g=1}^{N_f}\left( 
(\chi_{A\,f}^\dagger\chi_{A\,g})Q^{gf}+ 
(\chi_{B\,f}^\dagger\chi_{B\,g})(Q^\dagger)^{gf}\right)} 
\mbox{e}^{-N\qq^2\Tr \, Q^\dagger Q}  
\nn\\ 
&&\ \label{HStrafo} 
\eea 
where we have integrated out the matrices $\Phi$, performed the trace  
and made a Hubbard-Stratonovich transformation. The auxiliary matrices $Q$  
are now quadratic of size $N_f\times N_f$. 
The integrations over the Grassmann vectors $\chi_A$ and $\chi_B$  
in eqs. (\ref{step1}) and (\ref{HStrafo}) are now all Gaussian, leading to  
the following determinant 
\bea 
{\cal Z}^{(N_f,\nu)}_I(\mu;M)&\sim& 
\int dQ dQ^\dagger \det\left( 
\begin{array}{ll} 
(M+Q)^T\otimes\one_N                    & \mu\Sigma_3\otimes \tilde{\one}_N\\ 
\mu\Sigma_3\otimes \tilde{\one}_N^\dagger&  
(M+Q^\dagger)^T\otimes\one_{N+\nu}\\ 
\end{array} 
\right) 
\mbox{e}^{-N\qq^2\Tr \, Q^\dagger Q} 
\nn\\ 
&=& \int dQ dQ^\dagger  
\det[M+Q^\dagger]^\nu  
\label{Z1qqQ}\\ 
&&\times\det\left[(M+Q^\dagger)(M+Q) 
-\mu^2(M+Q^\dagger)\Sigma_3(M+Q^\dagger)^{-1}\Sigma_3\right]^N 
\mbox{e}^{-N\qq^2\Tr \, Q^\dagger Q}. 
\nn 
\eea 
Here we have introduced the mass matrix $M=$ diag$(\om_1,\ldots,\om_{N_f})$ 
and the generalized Pauli matrix of size $N_f\times N_f$ 
\be 
\Sigma_3\ \equiv\ \left( 
\begin{array}{cc} 
\one_m & 0\\ 
0 & -\one_n\\ 
\end{array} 
\right) . 
\label{sig} 
\ee 
Furthermore we made use of  
the following property of determinants for invertible matrices $D$ 
\be 
\det\left( 
\begin{array}{cc} 
A & B\\ 
C & D\\ 
\end{array} 
\right)\ =\ \det[D]\ \det[A\ -\ B\,D^{-1}C] \ . 
\ee 
For real valued masses, $M=M^\dagger$, we can further simplify 
eq. (\ref{Z1qqQ}) by shifting $Q+M\to Q$, 
\bea 
{\cal Z}^{(N_f,\nu)}_I(\mu;M)&=& \mbox{e}^{-N\qq^2\Tr\, M^2}  
\int dQ dQ^\dagger  
\det[Q^\dagger]^\nu \det\left[Q^\dagger Q-\mu^2Q^\dagger 
\Sigma_3Q^{\dagger\,-1}\Sigma_3\right]^N 
\nn\\ 
&&\ \ \ \ \ \ \ \ \ \ \ \ \ \ \ \ \ \ \ \times 
\mbox{e}^{-N\qq^2\Tr \left(Q^\dagger Q\ -\ M(Q+Q^\dagger)\right)}. 
\label{Z1Q} 
\eea 
In the absence of conjugate anti-quarks, $\Sigma_3=\one_{N_f}$, the second 
determinant further simplifies, leaving $\mu^2\one_{N_f}$ only. 
In order to arrive at a unitary group integral  
we introduce the ``polar decomposition'' 
$Q=U\,R$, where $U^\dagger=U^{-1}$ is an $N_f\times N_f$ unitary matrix  
and $R^\dagger=R$ is hermitian and positive. The Jacobian for the  
transformation is derived in the following appendix. 
An alternative is the Schur decomposition of the matrix $Q$  
used in \cite{A03} which  
quickly becomes cumbersome for more than $N_f=3$ flavors.  
The advantage of our present method is  
that we can conveniently exploit the saddle point approximation 
for the integrals over positive eigenvalues  
$r_i$ of the matrix $R$.

\sect{Jacobian for the polar decomposition}\label{polarJac} 
 
A classical result of linear algebra states that any $N_f\times N_f$ complex 
matrix $Q$ can be written as $Q=U\,R$ where $U^\dagger=U^{-1}$ is an 
$N_f \times N_f$ unitary matrix and $R^\dagger=R$ is $N_f\times N_f$ hermitian 
positive matrix. That is the matrix-generalization of the ``polar 
decomposition'' of a complex number $q=r e^{i \theta}$ Clearly enough  
the number of real degrees of freedom of the matrix $Q$ (i.e. $2N_f^2$)  
matches with the total number of real degrees of freedom of $U$ 
(i.e. $N_f^2$) and $R$ (i.e. $N_f^2$).  Under this matrix change of 
variables, the integration measure produces a Jacobian: 
\be 
\label{polartrans} 
dQ \, dQ^{\dagger}=J(R) \, d \mu(U) \, dR \, , 
\ee 
where $d \mu(U)$ is the Haar measure on the unitary group, $dR$ is 
the measure on the space of Hermitian positive matrices and $J(R)$ is 
the Jacobian that we are going to calculate in this 
appendix.  For calculating the Jacobian\footnote{For general methods of
evaluating Jacobians, see \cite{Mathai} \cite{Hua}.}, we 
differentiate: 
\be 
\label{polardiff} 
dQ        = dU \, R +U dR  \, . 
\ee 
As $R$ is an Hermitian matrix, it can be diagonalized by an unitary 
matrix $V$: $R=V \hat{r} V^{-1}$ with $\hat{r}=$ diag$(r_1,\ldots,r_n)$  
having positive entries. If 
we left-multiply eq. (\ref{polardiff}) by $V^{-1} U^{-1}$ and 
right-multiply by $V$, we get 
\be 
\label{polardiff2} 
V^{-1} U^{-1} \, dQ  \, V=  
V^{-1} U^{-1} \, dU \, V \, r  + V^{-1} \, dR \, V  \, . 
\ee 
This equation can be written as  
$d\tilde{Q}= d\tilde{U} \, r + d\tilde{R}$  
with  
$d\tilde{Q} \equiv V^{-1} U^{-1} \, dQ \, V$, 
$d\tilde{U} \equiv V^{-1} U^{-1} \, dU \, V $  
and  
$d \tilde{R} \equiv V^{-1} \, dR \, V $.   
The matrix $d\tilde{U}$ is anti-hermitian, since 
$(d\tilde{U})^{\dagger}= V^{\dagger} (dU)^{\dagger} U V= -V^{-1} 
U^{-1} \, dU \, V=- d\tilde{U} $  
(we used $dU^{\dagger} U=- U^{-1} \, dU$).   
Moreover, the matrix $d \tilde{R}$ is hermitian, since $d 
\tilde{R}^\dagger=(V^{-1} \, dR \, V )^{\dagger}= \tilde{R}$.  
Therefore, the relation between the differentials of the independent complex 
variables is: 
\be 
\label{polarcplx} 
\left\{ 
\begin{array}{ll} 
(d\tilde{Q})_{ii}= d\tilde{U}_{ii} \, r_i  + d \tilde{R}_{ii}    
        & i=1,\ldots,N_f  \\ 
(d\tilde{Q})_{ij}= d\tilde{U}_{ij} \, r_j  + d \tilde{R}_{ij}              \\ 
(d\tilde{Q})_{ji}= d\tilde{U}^{*}_{ij} \, r_i  + d \tilde{R}^{*}_{ij}    
&  i<j   
\end{array} 
\right. . 
\ee 
Let us introduce a real parameterization as follows: 
\bea 
(d\tilde{Q})_{ij}&=& dx_{ij}+i dy_{ij} \, , \quad 
(d\tilde{R})_{ij}=  
\left\{ 
\begin{array}{ll} 
(d\tilde{R})_{ij}= dp_{ij}+i dq_{ij} &  i<j \, , \\ 
(d\tilde{R})_{ii}= dp_{ii}  &  i=1,\ldots,N_f   
\end{array} 
\right. \nn \\ 
(d\tilde{U})_{ij} &=&  
\left\{ 
\begin{array}{ll} 
(d\tilde{U})_{ij}= dv_{ij}+i dw_{ij} &  i<j \, , \\ 
(d\tilde{U})_{ii}= i dw_{ii}  &  i=1,\ldots,N_f  
\end{array} 
\right. \nn  
\, . 
\eea 
By matching the real and imaginary parts we have: 
\begin{enumerate} 
\item[1)] $dx_{ii}=dp_{ii}$ and $dy_{ii}=r_i dw_{ii}$, for $i=1,\ldots,N_f$.  
The corresponding Jacobian 
is $J_1 = \prod_{i=1}^{N_f} r_i$. 
\item[2)] $dx_{ij}=dv_{ij} r_j+dp_{ij}$ and $dx_{ji}=-dv_{ij} r_i + dp_{ij} $  
for any pair $i<j$. The corresponding 
Jacobian is $(J_2)_{ij}= \det \frac{\partial (x_{ij},x_{ji} )} 
{\partial (v_{ij},p_{ij})}=\det  
\left| 
\begin{array}{cc} 
r_j & 1 \\ 
-r_i & 1  
\end{array} 
\right| 
=(r_i+r_j)$, and therefore $J_2 = \prod_{i<j}^{N_f} (r_i+r_j)$. 
\item[3)] $dy_{ij}=dw_{ij} r_j+dq_{ij}$ and $dy_{ji}=dw_{ij} r_i - dq_{ij} $  
for any pair $i<j$. The corresponding 
Jacobian is $J_3= \prod_{i<j}^{N_f} \det  
\left| 
\begin{array}{cc} 
r_j & 1 \\ 
r_i & -1   
\end{array} 
\right| 
=\prod_{i<j}^{N_f}(r_i+r_j)$. 
\end{enumerate} 
Therefore the full Jacobian is $J= \prod_{i=1}^{N_f} r_i \,  
\prod_{i<j}^{N_f}(r_i+r_j)^2$.  Going back to the original variables 
$d\tilde{U} \to dU$, $d\tilde{R} \to dR$ and $d\tilde{Q} \to dQ$ does 
not produce any additional factor in the final Jacobian (up to an 
overall sign) as they differ only by unitary transformations. We thus 
obtain the final result 
\bea 
dQ \, dQ^{\dagger}&=&\prod_{i=1}^{N_f} r_i \prod_{i<j}^{N_f}(r_i+r_j)^2 \,  
d \mu(U) \, dR \nn \\ 
&\propto& \prod_{i=1}^{N_f} r_i \prod_{i<j}^{N_f}(r_i+r_j)^2 \prod_{i<j}^{N_f}
(r_i-r_j)^2 \, d \mu(U) d \mu(V) \, \prod_{i}^{N_f}  dr_i \nn \\ 
&=& \prod_{i=1}^{N_f} r_i \, dr_i \prod_{i<j}^{N_f}=(r_i^2-r_j^2)^2  
d \mu(U) d \mu(V) \ , 
\label{finaljacob}  
\eea 
where we used the standard fact that $R=V \hat{r} V^{-1}$ implies   
$dR \propto \prod_{i<j}^{N_f}(r_i-r_j)^2 \prod_{i}^{N_f}  dr_i \, d \mu(V)$. 
We have thus shown eq. (\ref{finaljacob}) to be the expression for  
the Jacobian for the polar decomposition $Q= U\,  R$.

\end{appendix}


\indent


\begin{thebibliography}{99} 
 
\bibitem{SV93} E.V. Shuryak and J.J.M. Verbaarschot, { Nucl. Phys.} {\bf A560} 
(1993) 306 [hep-th/9212088]. 
 
\bibitem{VW} J.J.M.~Verbaarschot and T.~Wettig, 
{ Ann.\ Rev.\ Nucl.\ Part.\ Sci.} {\bf 50} (2000) 343 [hep-ph/0003017]. 
 
\bibitem{JSV} A.D. Jackson, M.K. Sener and J.J.M.~Verbaarschot, 
Phys. Lett. {\bf B387} (1996) 355 [hep-th/9605183].   
 
\bibitem{DOTV} 
J.C.~Osborn, D.~Toublan and J.J.M. ~Verbaarschot, 
Nucl.\ Phys.\ B {\bf 540} (1999) 317 [hep-th/9806110]; 
P.H.~Damgaard, J.C.~Osborn, D.~Toublan and J.J.M.~Verbaarschot, 
Nucl.\ Phys.\ B {\bf 547} (1999) 305 [hep-th/9811212]. 
 
\bibitem{TV1} 
D.~Toublan and J.~J.~M.~Verbaarschot, 
Nucl.\ Phys.\ B {\bf 603} (2001) 343 [hep-th/0012144]. 
 
\bibitem{SV03} 
K.~Splittorff  and J.J.M.~Verbaarschot, 
hep-th/0310271. 
 
\bibitem{ADp}G.~Akemann and P.H.~Damgaard, Phys. Lett. {\bf B583} (2004) 199   
[hep-th/0311171].   
 
\bibitem{Steph}M.A. Stephanov, Phys. Rev. Lett. {\bf 76} (1996) 
4472 [hep-th/9604003]. 
 
\bibitem{HJSSV}M.A. Halasz, A.D. Jackson, R.E. Shrock,  
M.A. Stephanov and J.J.M. Verbaarschot,  
Phys. Rev. {\bf D58} (1998) 096007 [hep-ph/9804290]. 
 
\bibitem{KTV} B.~Klein, D.~Toublan and J.J.M.~Verbaarschot, 
Phys. Rev. {\bf D68} (2003) 014009 [hep-ph/0301143]. 
 
\bibitem{TV} D.~Toublan and J.J.M.~Verbaarschot, 
Int. J. Mod. Phys. {\bf B15} (2001) 1404 [hep-th/0001110]. 
 
\bibitem{HJV}M.A. Halasz, A.D. Jackson and J.J.M. Verbaarschot,  
Phys. Rev. {\bf D56} (1997) 5140 [hep-lat/9703006]. 
 
\bibitem{HOSV}M.A. Halasz, J.C.~Osborn, 
M.A. Stephanov and J.J.M. Verbaarschot,  
Phys. Rev. {\bf D61} (2000) 076005 [hep-lat/9908018]. 
 
\bibitem{AANV} J. Ambj\o rn, K.N. Anagnostopoulos, J. Nishimura, and 
J.J.M. Verbaarschot, JHEP 0210 (2002) 062 [hep-lat/0208025]. 
 
\bibitem{A03}  G. Akemann, 
Acta Phys. Polon. {\bf B34} (2003) 4653 [hep-th/0307116]. 
 
\bibitem{FKS}Y.V. Fyodorov, B.A. Khoruzhenko and H.-J. Sommers, 
Phys. Lett. {\bf A226} (1997) 46  [cond-mat/9606173];  
Phys. Rev. Lett. {\bf 79} (1997) 557  [cond-mat/9703152]. 
 
\bibitem{FS}Y.V. Fyodorov, and H.-J. Sommers, 
{ J. Phys. {\bf A}: Math. Gen.} {\bf 36} (2003) 3303 [nlin.CD/0207051]. 
 
\bibitem{A02} G. Akemann, { Phys. Rev. Lett.} {\bf 89}, 072002 (2002) 
[hep-th/0204068];  
{ J. Phys. {\bf A}: Math. Gen.} {\bf 36} (2003) 3363 [hep-th/0204246]. 
 
\bibitem{AWII} G. Akemann and T. Wettig, 
Phys. Rev. Lett. {\bf 92} (2002) 102002 [hep-lat/0308003]; 
Nucl. Phys. {\bf B129-130c} Proc. Suppl. (2004) 527 [hep-lat/0309037]. 

\bibitem{Kanzieper} E.~Kanzieper, Phys.\ Rev.\ Lett.\  {\bf 89} (2002), 250201
[cond-mat/0207745]. 
 
\bibitem{ADMN} G.~Akemann, P.~H.~Damgaard, U.~Magnea and S.~Nishigaki, 
Nucl.\ Phys.\ B {\bf 487} (1997) 721 
[hep-th/9609174]. 
 
\bibitem{DN} P.H. Damgaard and S.M. Nishigaki, Nucl. Phys. {\bf B518} (1998) 
495; Phys. Rev. {\bf D57} (1998) 5299-5302 [hep-th/9711096]. 
 
\bibitem{A02II}G. Akemann, Phys. Lett. {\bf B547} (2002) 100 [hep-th/0206086]. 
\bibitem{KS} J.P. Keating and N. Snaith, Commun. Math. Phys. {\bf 214}  
(2000) 57.  
 
\bibitem{AS} A.~Andreev and B.D.~Simons,
Phys.\ Rev.\ Lett.\  {\bf 75} (1995) 2304. 
 
\bibitem{F} Y.V. Fyodorov, Nucl. Phys. {\bf B621} [PM] 
(2002) 643 [math-ph/0106006]. 
 
\bibitem{BH} E. Br\'ezin and S. Hikami,  
Commun. Math. Phys. {\bf 214} (2000) 111 [math-ph/9910005]. 
 
\bibitem{MN} M.L. Mehta and J.-M. Normand, J. Phys. {\bf A}:  
Math. Gen.  {\bf 34} (2001) 1 [cond-mat/0101469]. 
 
\bibitem{FW}  P.J. Forrester and N.S. Witte, Commun. Math. Phys. 219  
(2001) 357 [math-ph/0103025]. 
 
\bibitem{FS1} Y.V. Fyodorov and E. Strahov, Nucl. Phys. {\bf B647} [PM] 
(2002) 581 [hep-th/0205215]. 
 
\bibitem{SV02} K.~Splittorff and J.J.M.~Verbaarschot, 
Phys.\ Rev.\ Lett.\  {\bf 90} (2003) 041601 [cond-mat/0209594]. 
 
\bibitem{FA} Y.V.~Fyodorov and G.~Akemann, 
JETP Lett.\  {\bf 77} (2003) 438 
[Pisma Zh.\ Eksp.\ Teor.\ Fiz.\  {\bf 77} (2003) 513] 
[cond-mat/0210647]. 
 
\bibitem{SF2} E. Strahov and Y.V. Fyodorov, Commun. Math. Phys.  
{\bf 241} (2003) 343 [math-ph/0210010]. 
 
\bibitem{FS3} Y.V. Fyodorov and E. Strahov,  
{ J. Phys. {\bf A}: Math. Gen.} {\bf 36} (2003) 3203 [math-ph/0204051]. 
 
\bibitem{BDS} J. Baik, P. Deift and E. Strahov, 
J. Math. Phys. {\bf 44} (2003) 3657 [math-ph/0304016].  
 
\bibitem{AF} G. Akemann and Y.V. Fyodorov, Nucl. Phys. {\bf B664} [PM] (2003) 
457 [hep-th/0304095]. 
 
\bibitem{V} M. Vanlessen, math-ph/0306078. 
 
\bibitem{A01}G. Akemann, Phys. Rev. {\bf D64} (2001) 114021 [hep-th/0106053]. 
 
\bibitem{AVIII} G. Akemann and G. Vernizzi, 
{ Nucl. Phys.} {\bf B660}, 532 (2003) [hep-th/0212051]. 
 
\bibitem{Bergere:2003ht} M.C.~Bergere, hep-th/0311227. 
 
\bibitem{FK}Y.V~Fyodorov and B.A.~Khoruzhenko,
Phys.\ Rev.\ Lett.\  {\bf 83} (1999) 65 [cond-mat/9903043]. 

\bibitem{GW} T. Guhr and T. Wettig, 
J. Math. Phys. {\bf 37} (1996) 6395 [hep-th/9605110]. 
 
\bibitem{BRT} R.C. Brower, P.Rossi and C.-T. Tan, Phys. Rev. {\bf D23} (1981) 
942; A.B. Balantekin, Phys. Rev. {\bf D62} (2000) 085017 [hep-th/0007161]. 
 
\bibitem{AD} G. Akemann, P. H. Damgaard,  
Phys. Lett. {\bf B432} (1998) 390 [hep-th/9802174]; 
H.W. Braden, A. Mironov and A. Morozov, Phys. Lett. {\bf B514} (2001) 293 
[hep-th/0105169].  
 
\bibitem{SW} B. Schlittgen and T. Wettig, J. Phys. {\bf A}: Math. Gen. 
{\bf36} (2003) 3195 [math-ph/0209030]. 
 
\bibitem{KV} A.B.J. Kuijlaars and  M. Vanlessen,  
Com. Math. Phys. {\bf 243} (2003) 163 [math-ph/0305044]. 
 
\bibitem{PZJ} P. Zinn-Justin, Commun. Math. Phys. {\bf 194} (1998) 631  
[cond-mat/9705044]. 
 
\bibitem{James} J.C. Osborn, hep-th/0403131.  
 
\bibitem{Mathai} A.M. Mathai, ``Jacobians of matrix transformations and  
functions of matrix argument'', World Scientific Publishing  
Company, 1997. 
 
\bibitem{Hua} L.K. Hua, ``Harmonic analysis of functions of several complex 
variables in the classical domains'', American Mathematical Society, 
Providence, 1963.   
 
 
 
\end{thebibliography}
\end{document}